\documentclass[fleqn,usenatbib]{mnras}

\usepackage[T1]{fontenc}
\DeclareRobustCommand{\VAN}[3]{#2}
\let\VANthebibliography\thebibliography
\def\thebibliography{\DeclareRobustCommand{\VAN}[3]{##3}\VANthebibliography}
\usepackage{soul,color}

\usepackage{graphicx}	
\usepackage{amsmath}	
\usepackage{amssymb}	
\usepackage{newtxtext,newtxmath}
\usepackage{physics}
\usepackage{booktabs}
\usepackage{multirow}
\usepackage{lscape}
\usepackage{rotating}
\usepackage{epstopdf}
\usepackage{soul}
\usepackage{float}

\newcommand{\Msun}{\, \mathrm{M}_{\odot}}
\newcommand{\HBTHERONS}{{\tt HBT-HERONS}}
\newcommand{\HBTPLUS}{{\tt HBT+}}

\newcommand{\lowres}{\tt L400m7}
\newcommand{\midres}{\tt L200m6}
\newcommand{\highres}{\tt L025m5}

\title[Galaxy morphologies in COLIBRE]{The morphologies of present-day galaxies in the COLIBRE simulations}

\author[Victor J. Forouhar Moreno et al.]{Victor J. Forouhar Moreno$^{1}$\thanks{E-mail: forouhar@strw.leidenuniv.nl}, Joop Schaye$^{1}$, Matthieu Schaller$^{2,1}$, Aaron Ludlow$^{3}$, Robert J. McGibbon$^{1}$, \newauthor Alejandro Ben\'itez-Llambay$^{4}$, Evgenii Chaikin$^{5,1}$, Carlos S. Frenk$^{5}$, Filip Hu\v{s}ko$^{1}$, Sylvia Ploeckinger$^{6}$, \newauthor  Alexander J. Richings$^{7,8}$,  
James W. Trayford$^{9}$ \\
$^{1}$Leiden Observatory, Leiden University, PO Box 9513, 2300 RA Leiden, the Netherlands\\
$^{2}$Lorentz Institute for Theoretical Physics, Leiden University, PO Box 9506, 2300 RA Leiden, the Netherlands\\
$^{3}$International Centre for Radio Astronomy Research, University of Western Australia, 35 Stirling Highway, Crawley, Western Australia, 6009, Australia\\
$^{4}$Dipartimento di Fisica G. Occhialini, Università degli Studi di Milano Bicocca, Piazza della Scienza, 3 I-20126 Milano MI, Italy \\
$^{5}$Institute for Computational Cosmology, Department of Physics, University of Durham, South Road, Durham, DH1 3LE, UK\\
$^{6}$Department of Astrophysics, University of Vienna, Türkenschanzstrasse 17, A-1180 Vienna, Austria \\
$^{7}$Centre for Data Science, Artificial Intelligence and Modelling, University of Hull, Cottingham Road, Hull, HU6 7RX, UK\\
$^{8}$E. A. Milne Centre for Astrophysics, University of Hull, Cottingham Road, Hull, HU6 7RX, UK\\
$^{9}$Institute of Cosmology and Gravitation, University of Portsmouth, Dennis Sciama Building, Burnaby Road, Portsmouth PO1 3FX, UK
}

\date{Accepted XXX. Received YYY; in original form ZZZ}

\pubyear{2025}

\begin{document}
\label{firstpage}
\pagerange{\pageref{firstpage}--\pageref{lastpage}}
\maketitle 

\begin{abstract}
The diversity of galaxy morphologies and their relations with galaxy and halo properties is fundamental to understanding galaxy formation. Cosmological simulations of representative volumes can help disentangle the origin of observed correlations, but most suffer from two main limitations that affect morphologies: an over-pressurised interstellar medium and spurious interactions between stellar and dark matter particles. We present an overview of galaxy morphologies in the COLIBRE simulations, which address these limitations and reproduce many observed galaxy scaling relations. To quantify galaxy morphology, we use four (strongly-correlated) theory-space metrics, three kinematic and one spatial. We explore how different choices and limitations affect these indicators, including luminosity- versus mass-weighting, aperture size and shot noise. Overall, we find good convergence in present-day morphologies across two orders of magnitude in mass resolution. COLIBRE  predicts that kinematic morphology correlates strongly with stellar mass and colour, and that galaxies with stellar masses of $\approx(1-2)\times 10^{10}\,\Msun$ tend to be the most rotationally-dominated. At fixed stellar mass, the morphology of central galaxies correlates weakly with the properties of their host halo. Morphology  correlates more strongly with internal galaxy properties, with more disky galaxies being more gas-rich, having higher star formation rates and exhibiting younger and more extended stellar populations. Other properties, like the mass of the most massive black hole, the fraction of stars that are accreted and stellar metallicity, also correlate with morphology, but with correlation strengths sensitive to the stellar mass of the galaxy and whether it is a central or satellite. 
\end{abstract}




\section{Introduction}\label{Section:introduction }

The spatial distribution of stars within galaxies, itself connected to their kinematics \citep[e.g.][]{Binney.1978, Binney.2008, Cappellari.2008}, comes in a variety of shapes, such as discs, ellipticals, and more irregular configurations \citep[e.g.][]{Hubble.1926, deVaucouleurs.1959, Sandage.1961}. Since stars are essentially collisionless on galactic scales, their spatial and kinematic distributions at any given time reflect the orbital properties of the gas from which they formed, and any subsequent evolution they experienced through gravitational interactions. Studying the origin of the spatial and kinematic morphology of galaxies, and its correlation with other galaxy properties, is an essential part of understanding how galaxies form and evolve.

Several correlations between morphology and other internal galaxy properties are well known \citep[e.g.][]{Baldry.2004, Wuyts.2011, Driver.2022}. Elliptical galaxies tend to be red, quenched, gas-poor and dominate the galaxy number density above $M_{*} \approx 10^{11}\,\Msun$. On the other hand, disc galaxies are generally blue, star-forming, gas-rich and dominate the galaxy number density between $\approx 10^{10}-10^{11}\,\Msun$. Below this stellar mass range, galaxies tend to be more irregularly-shaped, although there is substantial morphological variety. There are in addition several secondary correlations that indicate that external factors play an important role in morphological transformations, like the number of neighbouring galaxies and the distance to a more massive galaxy \citep[e.g.][]{Oemler.1974, Dressler.1980, Postman.1984, Li.2006, Skibba.2009}.

Despite the knowledge of the above correlations, establishing the connection between the physics of galaxy formation and morphology from a purely observational point of view is not trivial. Galaxy properties can only be measured at a fixed moment in time, and so identifying the origins of observed correlations inherently requires making assumptions about the past evolution of galaxies. In this respect, cosmological hydrodynamical simulations play a key role by providing a view of the whole evolutionary history of any given galaxy, reducing the need to make strong assumptions to establish cause and effect.

Nonetheless, a basic question needs to be asked before using simulations to help interpret the data, which is whether the properties of simulated galaxies are realistic. The design of subgrid models for unresolved baryonic processes relevant to galaxy formation involves making choices, some of which lead to the introduction of free parameters. The subgrid modules are therefore `calibrated', either explicitly or implicitly, by comparing the predictions of the model to a chosen subset of observations, which can include the stellar mass function \citep[e.g.][]{Genel.2014}, the size - stellar mass relation \citep[e.g.][]{Crain.2015}, the black hole mass - stellar mass relation \citep[e.g.][]{Booth.2009}, gas fractions in clusters \citep[e.g.][]{McCarthy.2017, Kugel.2023}, the intracluster gas density profile \citep[e.g.][]{Frontiere.2025} and the reionization history \citep[e.g.][]{Pawlik.2017}. A clear disadvantage of the calibration procedure is that the relations used to anchor the subgrid choices no longer constitute true predictions. Nonetheless, alongside other developments in the past two decades, this approach has taken us from unrealistically compact and massive galaxy populations to galaxies that reproduce many observed galaxy scaling relations even beyond calibrated ones (see \citealt{Crain.2023} for a recent review). A common factor across the aforementioned simulations is that none use morphology to directly calibrate the subgrid model parameters\footnote{Calibrating the size - stellar mass relation does indirectly affect galaxy morphology, because bulges are more compact than discs.}. It is therefore interesting to examine what simulations predict, since they may agree in overall stellar mass with observations, but differ greatly in its phase-space distribution within galaxies. Indeed, galaxy morphologies are sensitive to the details of the subgrid model physics \citep[e.g.][]{Okamoto.2005,Scannapieco.2012, Zhang.2024, Celiz.2025}.

Comparing observed and predicted galaxy morphologies reveals a mix of successes and outstanding challenges. Some simulations produce over-massive discs in low- \citep[][]{Bottrell.2017}, intermediate- \citep[][]{Dickinson.2018} and high-mass \citep[][]{Huertas-Company.2019, Jung.2022} galaxies, but others are able to qualitatively reproduce trends with stellar mass \citep{Tacchella.2019}, colour \citep[][]{Correa.2017} and environment \citep[][]{Pfeffer.2023}. However, even simulations capable of reproducing basic trends reveal a more nuanced view under more detailed scrutiny, like overly-round high- \citep[][]{deGraaff.2022} and low-mass \citep[][]{Klein.2025, Benavides.2025} galaxies, a mismatch in stellar kinematics \citep[][]{Lagos.2018, vandeSande.2019}, and too-asymmetric galaxies \citep[][]{Bignone.2020}.

A complicating factor is that existing cosmological simulations of representative volumes often suffer from two severe limitations when it comes to predicting the morphologies of galaxies. First, with the exception of Romulus \citep[][]{Tremmel.2017}, NewHorizon \citep[][]{Dubois.2021} and FIREbox \citep[][]{Feldmann.2023}, gas is artificially prevented from cooling down below a temperature floor of $\approx10^{4}\,\mathrm{K}$ and reaching high densities. In practice, the artificial over-pressurisation of the interstellar medium limits the spatial resolution, and stars are therefore born from gas discs that are too thick, propagating these numerical artefacts into their resulting morphologies.

The second limitation is that cosmological simulations contain finite numbers of baryonic and dark matter (DM) particles. Most cosmological simulations, with the exception of Romulus \citep[][]{Tremmel.2017}, adopt equal numbers for both species, such that the DM-to-baryonic particle mass ratio is $\Omega_{\rm DM} /\Omega_{\rm bar} \approx 5$. This is troublesome because the number of particles affects two forms of energy exchange between the DM and stellar components as the system evolves towards energy equipartition through gravitational scattering. The first form of energy transfer originates from unequal particle masses, which enhance two-body energy transfer from the heavier (DM) to the lighter (stellar) species \citep[e.g.  mass segregation;][]{Sellwood.2013, Ludlow.2019}. The second energy exchange, which is typically dominant and present even if the particle mass ratio is unity, arises because the stellar component is dynamically colder than the surrounding DM. Thus, the DM heats the stellar component at a rate that depends on the halo relaxation time, which is set by the number of dark matter particles. Both mechanisms transfer energy from DM to stellar particles and convert ordered rotation into random motion, increasing the velocity dispersions and sizes of galaxies, and the scale heights of discs \citep[][]{{Sellwood.2013, Revaz.2018, Wilkinson.2023, Ludlow.2023}}. In the real Universe, these effects are negligible because the DM particle mass is microscopic and the corresponding halo relaxation time is exceedingly long.

With this context in mind, the new COLIBRE suite of simulations \citep[][]{Schaye.2025, Chaikin.2025} represents a step forward towards predicting and studying galaxy morphologies in a cosmologically-representative setting. COLIBRE does not use an artificially over-pressurised interstellar medium, allowing stars to form from gas with higher densities and lower temperatures than simulations of comparable cosmic volume. Additionally, dark matter particles are super-sampled (4 DM particles per baryonic particle), such that their mass relative to baryonic particles is close to unity (suppressing energy transfer by mass segregation) and so that the number of DM particles per halo increases (prolonging their relaxation times). COLIBRE also introduces many other improvements compared to previous generation large-volume hydrodynamical simulations. It uses a refined model of feedback, allowing feedback events to be better sampled, and providing two separate models of active galactic nuclei feedback (purely thermal \textit{versus} thermal plus kinetic jets). Dust creation, growth and destruction are computed on-the-fly, which couples to a non-equilibrium chemistry network and cooling tables. 

After calibrating the relevant subgrid parameters, COLIBRE is able to reproduce a variety of observed galaxy properties beyond its calibration targets. At $z = 0$ , the specific star formation rate of active galaxies, the X-ray emission from the circumgalactic medium, gas and stellar metallicities, and the dust, \ion{H}{I} and $\mathrm{H}_{2}$ content of galaxies are generally in agreement with observations \citep[e.g.][]{Schaye.2025}. COLIBRE has also been shown to reproduce the stellar mass function of galaxies up to the highest redshifts where there is observational data \citep[][]{Chaikin.2025b}, the observed angular momenta and sizes over the full stellar mass range sampled by COLIBRE for a wide range of characteristic radii and for both early- and late-type galaxies, out to $z\approx 1$ \citep[][]{Ludlow.2026}, as well as the median trends and scatter of the spatially resolved Kennicutt-Schmidt relation relation for \ion{H}{I} and $\mathrm{H}_{2}$ up to $z = 5$ \citep[][]{Lagos.2025}. 

In this work, we examine the $z = 0$ morphologies of galaxies predicted by COLIBRE. Since this paper is the first to do so, we choose to quantify morphologies in `theory-space' using four different morphology metrics: one spatial (flattening of the stellar distribution) and three kinematic (the spheroid mass fraction, the fraction of kinetic energy in co-rotation and the rotation-to-dispersion velocity ratio). This approach implies that we have full knowledge of the phase-space distribution of stars, without associated observational uncertainties (e.g. sample selection, projection effects, background contamination, instrumental effects). Our approach allows us to explore how certain operational choices affect the measurement of intrinsic galaxy morphologies, such as the choice of morphology metric, whether averages are mass-weighted or luminosity-weighted and how stellar particles are selected. The caveat of this choice is that it severely limits comparisons to observational data. Hence, we postpone a comparison to observations for future work, in which the relevant observational effects are modelled to consistently compare simulations to observations.

We structure this paper in the following manner. We first expand on the main features of the COLIBRE subgrid model and how galaxies are identified within the simulations in \S\ref{Section:Simulations}. Next, in \S\ref{Halo_property_calculation} we explain how we compute halo and galaxy properties. The definitions for our selected morphology metrics, together with the impact of various operational choices (e.g. aperture sizes, luminosity-weighting), are discussed in \S\ref{Section:MorphologicalCalculations}. We quantify the minimum number of particles required for accurate morphologies and the convergence across mass resolutions in \S\ref{Section:ConvergeMorphology}. The primary results are presented in sections \S\ref{Section:MorphologicalCorrelations} through \S\ref{Section:CorrelationMorphologyOtherGalaxyProperties}, in which we correlate various morphology parameters with the environment, host halo properties, and internal galaxy characteristics, such as stellar mass, colour, size, and black hole mass.

\section{Methods}\label{Section:Methods}

In this section we first outline the main features of the COLIBRE galaxy formation model, how haloes and galaxies are found within the simulations and which subset of the simulation suite we use in this study (\S\ref{Section:Simulations}). We then explain how we measure the properties of haloes and galaxies found in the simulations (\S\ref{Halo_property_calculation}) and how we match haloes between dark-matter-only (DMO) and hydrodynamical simulations (\S\ref{Section:SubhaloMatching}).

\subsection{Simulations}\label{Section:Simulations}

The COLIBRE simulations\footnote{\hyperlink{https://www.colibre-simulations.org}{https://www.colibre-simulations.org}} \citep[][]{Schaye.2025, Chaikin.2025} are a set of cosmological hydrodynamical and DMO simulations. All simulations were run with the {\tt SWIFT} code \citep{Schaller.2024}, using the {\tt SPHENIX} implementation of smoothed particle hydrodynamics \citep{Borrow.2022}. The initial conditions were generated by {\tt monofonic} \citep[][]{Hahn.2021, Michaux.2021} using the cosmological parameters taken from the 3x2pt plus external constraints from DES Y3 \citep[$\Omega_{\rm CDM} = 0.256011$, $\Omega_{\rm b} = 0.0486$, $\Omega_{\Lambda} = 0.693922$ and $H_{0} = 68.1\,\mathrm{km}\,\mathrm{s}^{-1}\,\mathrm{Mpc}^{-1}$;][]{Abbott.2022}.

Contrary to the previous generation of simulations that target representative cosmic volumes,  gas in COLIBRE can cool down to $\approx 10\,\mathrm{K}$. Its cooling and heating rates are determined by {\tt HYBRID-CHIMES} \citep[][]{Ploeckinger.2025}, using the non-equilibrium {\tt CHIMES} chemistry network \citep[][]{Richings.2014a, Richings.2014b}. The rates in {\tt HYBRID-CHIMES} are computed using the non-equilibrium abundances of hydrogen, helium and their free-electrons, take into account the metals that dominate the radiative cooling (C, N, O, Ne, Mg, Si, S, Ca,
and Fe), the gas dust content and dust shielding, as well as the cosmic microwave background, an interstellar radiation field and a time-evolving UV/X-ray background. 

Star formation proceeds as described in \citet[][]{Nobels.2024}. In short, gas particles become eligible to form stars if the gas within the SPH kernel is gravitationally unstable, taking into account the turbulent and thermal velocity dispersion. Eligible particles are converted stochastically into stellar particles following the \citet[][]{Schmidt.1959} law, with an efficiency per free-fall time of 1 per cent.

The model for metal enrichment includes stellar mass loss from asymptotic giant branch and massive stars, core-collapse supernovae, supernovae type Ia, neutron star mergers and collapsars \citep[][]{Correa.2026}. 
A model for the unresolved small-scale turbulent mixing of metals is detailed in \citet[][]{Correa.2026}. Dust creation, growth and destruction \citep[][]{Trayford.2025} is done on-the-fly, and couples self-consistently to the cooling and chemical reaction network.

Various channels of energy feedback are included in the simulations, from pre- and post-supernova stellar feedback to feedback from active galactic nuclei (AGN). Pre-supernova feedback includes stellar winds, radiation pressure and thermal energy sourced by \ion{H}{II} regions \citep[][]{BenitezLlambay.2025}. Supernova feedback is injected isotropically, with 10 per cent of energy in kinetic mode and the remainder in the form of a temperature jump whose magnitude depends on the local gas density \citep[][]{Chaikin.2023, Schaye.2025}. AGN feedback is sourced by black holes, which can accrete up to 100 times the Eddington accretion rate. The form in which AGN feedback energy is deposited, as well as the feedback efficiency of the black hole, depends on which of the two implemented AGN feedback modes is used. For the purely thermal mode, the radiative efficiency is fixed and particles are heated using a temperature jump that scales with the mass of the black hole particle \citep[][]{Booth.2009, Schaye.2025}. The hybrid AGN model includes a thermal and a bipolar kinetic jet mode, and the feedback efficiency depends on the black hole spin and its accretion rate, as described in \citet[][]{Husko.2025}.

\begin{table*}
\centering
\caption{Overview of the $z = 0$ COLIBRE simulations used for the main results of this work. From left to right, each column indicates the shorthand identifier for each simulation, the mean dark matter particle mass ($m_{\rm DM}$), the initial mean baryonic particle mass ($m_{\rm bar}$), the maximum physical gravitational softening length ($\epsilon_{\rm max}$), the comoving side-length of the box ($L_{\rm box}$), the AGN feedback mode and the number of total ($N_{\rm gal}$), central ($N_{\rm cen}$) and satellite galaxies ($N_{\rm sat}$) which contain at least 50 bound stellar particles within 50~kpc and hence are not significantly affected by shot noise (see \S\ref{convergence_sampling_size}). We use the largest common volume available to all three resolutions at $z= 0$ ($L_{\rm box} = 25\,\mathrm{Mpc}$) for convergence testing. The validation of methods to quantify galaxy morphology uses the {\midres} simulation. The main results combine the largest boxes available for each resolution.}
\begin{tabular}{@{}lllrrrrrr@{}}
\toprule
Identifier & $m_{\rm DM}\,[\mathrm{M}_{\odot}]$ & $m_{\rm bar}\,[\mathrm{M}_{\odot}]$ & $\epsilon_{\rm max}\,[\mathrm{kpc}]$ & $L_{\rm box}\,[\mathrm{Mpc}]$ & AGN mode & $N_{\rm gal}$ & $N_{\rm cen}$ & $N_{\rm sat}$      \\ \midrule
L400m7     & $1.94\times10^7$                   & $1.47\times10^7$                    & 1.40                                  & 400 & Thermal                           & 1230183  & 702453 & 527730\\
L400m7-DMO     & $1.84\times10^7$                   & -                    & 1.40                                  & 400 & -                          & -  & - & -\\
L200m7     & $1.94\times10^7$                   & $1.47\times10^7$                    & 1.40                                  & 200 & Thermal                          & 154765  & 88538 & 66227 \\
L200m7h     & $1.94\times10^7$                   & $1.47\times10^7$                    & 1.40                                  & 200  & Hybrid                         & 155466  & 87135 & 68331 \\
L025m7     & $1.94\times10^7$                   & $1.47\times10^7$                    & 1.40                                  & 25  & Thermal                          & 303   & 177 & 126                \\
L025m7-DMO     & $1.84\times10^7$                   & -                    & 1.40                                  & 25  & -                          & -   & - & -                \\ \midrule
L200m6     & $2.42\times10^6$                   & $1.84\times10^6$                    & 0.70                                  & 200  & Thermal                         & 367117  & 204854 & 162263 \\
L200m6-DMO     & $2.30\times10^6$                   & -                    & 0.70                                  & 200   & -                        & -  & - & - \\
L100m6     & $2.42\times10^6$                   & $1.84\times10^6$                    & 0.70                                  & 100 & Thermal                          & 45410  & 25715 & 19695 \\
L100m6h     & $2.42\times10^6$                   & $1.84\times10^6$                    & 0.70                                  & 100 & Hybrid                          & 44868  & 25462 & 19406 \\
L025m6     & $2.42\times10^6$                   & $1.84\times10^6$                    & 0.70                                  & 25 & Thermal                           & 748                  & 421 & 327\\
L025m6-DMO     & $2.30\times10^6$                   & -                    & 0.70                                  & 25 & -                           & -                  & - & -\\
L025m6-DMm7     & $9.68\times10^6$                   & $1.84\times10^6$                    & 0.70                                  & 25 & Thermal                           & 751                  & 425 & 326\\
\midrule
L025m5     & $3.03\times10^5$                   & $2.30\times10^5$                    & 0.35                                 & 25 & Thermal                            & 1761               & 995 & 766\\
L025m5h     & $3.03\times10^5$                   & $2.30\times10^5$                    & 0.35                                 & 25 & Hybrid                           & 1755               & 990 & 765\\
L025m5-DMO     & $2.87\times10^5$                   & -                    & 0.35                                 & 25 & -                           & -               & - & -\\ \bottomrule

\end{tabular}\label{Table:simulations_used}
\end{table*}

Four of the parameters of the COLIBRE subgrid model were calibrated to observed data using a mix of machine learning and manual adjustments \citep[][]{Chaikin.2025}. The calibration targets were the $z = 0$ stellar mass function inferred from GAMA \citep[][]{Driver.2022} and the $z < 0.05$ galaxy size-mass relation \citep[][]{Hardwick.2022}, for galaxies with stellar masses between $10^{9}\,\Msun$ and $10^{11.3}\,\Msun$. The black hole mass - stellar mass relation inferred from dynamical measurements of massive nearby galaxies \citep[][]{Graham.2023} was also used during calibration. Beyond the data it was calibrated against, COLIBRE has already been shown to reproduce a diverse set of observed galaxy population properties at low and high redshifts (e.g. \citealt{Schaye.2025, Chaikin.2025b, Lagos.2025,Ludlow.2026})

Three different resolutions are available in COLIBRE, each covering a range of cosmic volumes. In this study we use several boxes and resolutions in combination, so that we can explore morphology across a wide range of stellar masses, as well as quantify convergence and the effect of different operational choices. We list which simulations we use in Table~\ref{Table:simulations_used}, with the main results of this paper being primarily based on the analysis of the simulations with the fiducial, purely thermal AGN feedback. The present-day morphologies of galaxies in simulations that use hybrid AGN feedback are very similar to those in the thermal AGN feedback simulations (see Appendix \ref{Appendix:hybrid_vs_thermal_morphologies}). 

We identify haloes in the simulation via the Friends-of-Friends (FoF) algorithm using a linking length equal to 0.2 times the mean dark-matter-interparticle separation. Subhaloes and their associated galaxies are identified using the history-based subhalo finder {\HBTHERONS}\footnote{\hyperlink{https://github.com/SWIFTSIM/HBT-HERONS/}{https://github.com/SWIFTSIM/HBT-HERONS/}} \citep{Forouhar.2025}, an updated version of {\HBTPLUS} \citep{Han.2018} that improves the identification and tracking of subhaloes in hydrodynamical and DMO simulations. We use {\HBTHERONS} as our fiducial choice because of its low computational cost, its robust identification of subhaloes \citep[][]{Forouhar.2025} and robust merger trees \citep{Chandro.2025}. 

For this work, we use central galaxies from {\midres} to asses the effect of operational choices in quantifying galaxy morphologies (\S\ref{Section:MorphologicalCalculations}). The bulk of the main results (\S\ref{Section:results}) use the whole galaxy population, which is in certain cases further subdivided into central and satellite galaxies. Except for convergence tests, each numerical resolution is assigned to a given stellar mass range, chosen to balance the needs for high resolution and large-number statistics. The only exception to the use of the whole galaxy population is during the investigation of the relation between galaxy morphology and the properties of the host halo (\S\ref{Section:CorrelationMorphologyHaloProperties}), where we only use central galaxies because spherical overdensities are ill-defined for satellites.

\subsection{Halo and galaxy properties}\label{Halo_property_calculation}

We use the {\tt SOAP}\footnote{\hyperlink{https://github.com/SWIFTSIM/SOAP}{https://github.com/SWIFTSIM/SOAP}} (\textbf{S}pherical \textbf{O}verdensity and \textbf{A}perture \textbf{P}rocessor; \citealt{McGibbon.2025}) Python package to measure the properties of subhaloes and galaxies. {\tt SOAP} measures various subhalo and galaxy properties by loading the centres of subhaloes and the particles in the surrounding region. We use as the centre of haloes and subhaloes the position of the most bound particle, which can be of any type. 


We measure seven different properties for haloes using {\tt SOAP}. The virial mass ($M_{\rm 200c}$), concentration ($c_{\rm 200c}$), spin parameter ($\lambda_{\rm 200c}$) and two shape indicators (sphericity and triaxiality) are calculated using all particles (of any type) within a sphere with a mean enclosed density of 200 times the critical density of the Universe\footnote{Using iterative inertia tensors may result in the inclusion of particles beyond $R_{\rm200c}$, as the sphere is deformed in a volume-conserving manner} ($\rho_{\rm crit}$). The aperture of the sphere ($R_{\rm 200c}$) is found by locating the smallest radius at which the overdensity is less than 200. By definition, the virial mass is:
\begin{equation}
    M_{\rm 200c} = \dfrac{4\pi R^{3}_{\rm 200c}}{3}\times 200\rho_{\rm crit} \,.
\end{equation}
The concentration of the halo is estimated from the first moment ($\int \rho(r)rdr$) of the total matter density distribution assuming an NFW \citep[][]{Navarro.1997} profile \citep[][]{Wang.2024}. For the halo spin, we use the \citet[][]{Bullock.2001} definition:
\begin{equation}
    \lambda_{\rm200c} = \dfrac{|\vec{L}_{\rm 200c}|}{\sqrt{2}M_{\rm 200c}V_{\rm200c}R_{\rm 200c}}\, ,
\end{equation}
where $\vec{L}_{\rm 200c}$ is the total angular momentum of all particles within $R_{\rm 200c}$ and $V_{\rm 200c}$ is the halo virial velocity:
\begin{equation}
    V_{\rm 200c} = \sqrt{\dfrac{GM_{\rm 200c}}{R_{\rm 200c}}}.
\end{equation}
We measure the shape of the halo using the square root of the eigenvalues (i.e. the principle axes $a$, $b$ and $c$) of the mass-weighted inertia tensor:
\begin{equation}
    I_{jk} \equiv \dfrac{\sum_{i} m_{i}x_{j}x_{k}}{\sum_{i} m_{i}} \, ,
\label{Equation:inertia_tensor}
\end{equation}
where $j$ and $k$ are indices that indicate the spatial dimension being used. We use the convention that $a \geq b \geq c$, so that $b/a$ and $c/a$ correspond to the intermediate-to-major and minor-to-major axis ratios, respectively. The inertia tensor is computed iteratively, whereby the initial spherical aperture is progressively deformed in a volume-conserving manner (the product $a\times b\times c$ remains constant; \citealt[][]{Warren.1992}) into an ellipsoid that aligns with the current principle axes of the enclosed particle distribution. The ellipsoid is used at each step to select the particles from which the next inertia tensor is computed, and the iterations stop once the length of the major axis ($a$) has converged to within 0.01 per cent.  We use this iterative approach because enforcing a spherical aperture biases the measured shapes, which will be particularly important for the shapes of galaxies, as discussed \S\ref{Subsection:IterativeInertiaTensors}. The sphericity of the halo is defined as its minor-to-major axis ratio ($c_{\rm halo}/a_{\rm halo}$) and its triaxiality as $T_{\rm halo} \equiv (a^{2}_{\rm halo}-b^{2}_{\rm halo})/(a^{2}_{\rm halo}-c^{2}_{\rm halo})$. We use the subscript `halo' to distinguish the principle axes of haloes from those of galaxies. 

We also compute the maximum circular velocity ($V_{\rm max}$) of haloes. We do so because it is an alternative indicator of the potential well depth that is measured on intra-halo scales, and may therefore be more relevant than halo mass for the galaxies that haloes host at their centre. Contrary to the other halo properties, we only use bound particles to compute $V_{\rm max}$. The value assigned to $V_{\rm max}$ is the peak value of the circular velocity curve, which is measured at each particle position from the halo centre out to $R_{\rm 200c}$:
\begin{equation}
    V_{\rm max} = \sqrt{\dfrac{GM(\leq R_{\rm max})}{R_{\rm max}}}\,,
\end{equation}
where $R_{\rm max}$ is the radius at which the maximum circular velocity is reached. Note that we place all particles within a gravitational softening length ($\epsilon_{\rm max}$) from the halo centre at a distance of $r = \epsilon_{\rm max}$, ensuring that $R_{\rm max} \geq \epsilon_{\rm max}$ and preventing spuriously high $V_{\rm max}$ values from being measured.

The last halo property we consider is $D_{1,0.1}$, which is a dimensionless measurement of its environment that does not correlate significantly with halo mass and therefore removes confounding trends between mass and environment\footnote{More massive haloes contain more neighbouring galaxies above a given mass threshold than lower mass haloes, and would thus be characterised as residing in denser environments based on other commonly-used metrics} \citep[][]{Haas.2012}. We use the distance to the nearest halo whose $M_{\rm200c}$ is at least one tenth as large as the virial mass of the reference halo, normalised by the $R_{\rm 200c}$ of the reference halo. Note that using a virial mass threshold excludes nearby massive satellites from being possible nearest neighbours, as the convention used in the SOAP catalogues generated for COLIBRE means that satellites have $M_{\rm200c} = 0$. We tested a modification of this metric that did not exclude them by selecting on total bound mass rather than virial mass, but our results were insensitive to whether satellites were included or excluded.


For galaxy properties, we only consider stellar and gas particles that are bound to a galaxy, and define the spatial centre of a galaxy as the position of its most bound particle, which can be of any particle type. The systemic velocity of the galaxy is taken to be the centre-of-mass velocity of all of its bound stellar particles. Depending on the property being measured, we use two different types of spatial apertures. For the properties that quantify the morphology of a galaxy, which we motivate and discuss in \S\ref{Section:MorphologicalCalculations}, we use three times the half-stellar-mass radius\footnote{The half-mass stellar radius is the radius that encloses half of the total stellar mass bound to a galaxy, i.e. it can include stellar mass beyond 50~kpc.} ($3 R_{1/2}$). All other galaxy properties use a fixed physical aperture of 50~kpc, which is sufficiently large to encompass most of our galaxies, with the exception of those with $M_{*} \geq 10^{11}\,\Msun$ \citep{Chaikin.2025b}.

\subsection{Matching DMO and hydrodynamical haloes}\label{Section:SubhaloMatching}

As part of this study, we quantify how galaxy morphology correlates with the properties of host haloes in \S\ref{Section:CorrelationMorphologyHaloProperties}. However, the properties of haloes can change as a result of baryonic physics, meaning that the strength of correlations may vary depending on whether halo properties are measured in the hydrodynamical or DMO versions of the simulations. To study baryonic effects on halo structure and their corresponding effect on morphology-halo correlations, we measure halo properties in the hydrodynamical simulations, as well as for their matched DMO counterparts.

We match haloes between the DMO and hydrodynamical versions of the COLIBRE simulations in the following manner. For each central subhalo in simulation $i$, we collect all of the dark matter particles that are bound to it. We then find via particle ID matching which FoF groups in simulation $j$ contain these particles, establishing a match $m_{i\rightarrow j}$ with the central subhalo of the FoF group that contains the largest share. This process is done in reverse to provide a match $m_{j\rightarrow i}$. Only subhaloes which are matched bijectively across simulations, i.e. $m_{i\rightarrow j} = m_{j\rightarrow i}$, are kept. By only selecting central subhaloes when matching in either direction, we also ensure that the match is between subhaloes that are centrals in both simulations. This procedure results in 95.1 (93.2) per cent of central galaxies with $M_{*}\geq 10^{8}\,\Msun$ ($M_{*}\geq 10^{7}\,\Msun$) in the {\midres} box being bijectively matched between the DMO and hydrodynamical versions. The small fraction of unmatched haloes have few particles and are typically poorly resolved.

\section{Morphology metrics}\label{Section:MorphologicalCalculations}

We quantify the morphology using several mass-weighted spatial and kinematic metrics. All of them are measured in `theory-space', i.e. by leveraging the full information provided by the simulation and without associated observational uncertainties. Although we do provide a comparison between mass- and luminosity-weighted quantities in \S\ref{Appendix:LuminosityWeighting}, we defer a consistent comparison of virtual and real observations to future work.

All of the morphology indicators we measure use an aperture equal to three times the half-stellar-mass radius ($3R_{\rm 1/2}$). We scale the aperture with $R_{\rm 1/2}$ because a fixed physical aperture corresponds to a different fraction of a galaxy’s size, depending on its stellar mass and size. The factor of three is chosen because it is sufficiently small to reduce the influence of the kinematically hotter and more spherical diffuse stellar distribution around galaxies (i.e. stellar haloes and intracluster light), whilst being sufficiently large to encompass most of the stellar disc the galaxy may have. More details are provided in \S\ref{Appendix:ApertureSize}.

To quantify the spatial distribution of stars, we measure the flattening of the stellar distribution as quantified by the minor-to-major axis ratio of the principle axes of the mass-weighted inertia tensor. The inertia tensor is computed iteratively (as detailed in \S\ref{Halo_property_calculation}) because enforcing a spherical aperture biases the measured shapes and consequently lowers the fraction of thin galaxies, as we show in \S\ref{Subsection:IterativeInertiaTensors}.

All the other metrics we use rely on stellar kinematics, and hence require a reference velocity vector against which to compare the motion of stars. We choose to use the mass-weighted stellar angular momentum ($\vec{L}_{*}$) within $3R_{\rm 1/2}$ for the purpose of determining whether stars are rotating, and if so, whether they are counter- or co-rotating relative to $\vec{L}_{*}$. 

The first kinematic metric we use is the fraction of kinetic energy in ordered co-rotation \citep[][]{Correa.2017}:
\begin{equation}
\label{equation:kappa_corot}
\kappa_{\rm corot} \equiv  \dfrac{K_{\rm corot}}{K_{\rm total}} = \dfrac{\sum_{i}  m_{i} (L_{z,i}/ m_{i}R_{i})^{2}\theta_{\rm H}(L_{z,i})}{\sum_{i} m_{i}v^{2}_{i}} \, ,
\end{equation}
where $L_{z,i}$ is the angular momentum of the \textit{i}-th star particle projected onto $\vec{L}_{*}$ and $R_{i}$ is its distance to the axis of rotation defined by $\vec{L}_{*}$. The Heaviside step function $\theta_{\rm H}$ ensures that only particles that are co-rotating contribute to the sum in the numerator. This definition differs from the other commonly used rotational support metric proposed by \citet{Sales.2010}, which also includes in the numerator of Eq.~\ref{equation:kappa_corot} the contribution of counter-rotating particles. Hence, $\kappa_{\rm corot} \leq \kappa_{\rm rot}$.

The second metric is motivated by the fact that the distribution of orbital circularities ($\epsilon_{\rm circ}$\footnote{The orbital circularity of a star is the ratio of its angular momentum component parallel to the disc of the galaxy, relative to the total angular momentum of a circular orbit with the same binding energy as the star \citep[][]{Binney.2008}.}) of stars is related to the presence or lack of certain galaxy components, such as a disc or a bulge. Thus, quantifying how much stellar mass is contained within certain values of $\epsilon_{\rm circ}$ provides an estimate of the mass of the component associated with those circularities. In this work, we measure a `spheroid' mass for each galaxy by assuming that all stars with $\epsilon_{\rm circ} \leq 0$ are associated with it, and that the circularity distribution of the spheroid is symmetric about $\epsilon = 0$ \citep[e.g.][]{Abadi.2003}. We call this component the spheroid, as opposed to the bulge, because its mass may include both the bulge mass and a fraction of the stellar halo mass\footnote{We limit the contribution of the stellar halo by our choice of aperture size, but some amount of contamination is inevitable when using a purely spatial selection, as stellar haloes can extend into the main galaxy.}. As our definition of spheroid mass is twice the mass of the counter-rotating stars, we can measure the spheroid-to-total stellar mass ratio as:
\begin{equation}
    \Bigg(\dfrac{M_{\rm s}}{M_{\rm t}}\Bigg)_{\rm kin} = \dfrac{2\sum_{i} m_{i}[1-\theta_{\rm H}(L_{z,i})]}{\sum_{i} m_{i}}\;.
\label{Equation:spheroid_to_total}
\end{equation}
We use the subscript `kin' to remind the reader of its purely kinematic definition. Note that discs components that are counter-rotating relative to the total angular momentum of galaxies are classified as `spheroids' by this definition \citep[e.g.][]{Algorry.2014}, although their prevalence is low \citep[][]{Ebrova.2021, Bevacqua.2022}. 

The last kinematic metric we consider is the ratio of the azimuthal rotational velocity ($V_{\phi}$) and the one-dimensional velocity dispersion ($\sigma_{\rm 1D}$). To compute both quantities, we use a cylindrical coordinate system whose $z$-axis is aligned with $\vec{L}_{*}$. The rotational velocity of the galaxy is the mass-weighted azimuthal velocity of all bound stars within the chosen aperture:
\begin{equation}
    V_{\phi} = \dfrac{\sum_{i} m_{i}v_{\phi,i}}{\sum_{i} m_{i}} \, .
    \label{Equation:rotational_velocity}
\end{equation}
The one-dimensional velocity dispersion of the galaxy is derived from the quadrature sum of the velocity dispersion along each cylindrical direction, normalised by a factor of $\sqrt{3}$:
\begin{equation}
    \sigma_{1\mathrm{D}} = \dfrac{1}{\sqrt{3}}\sqrt{\sum_{j\in\{r,\phi,z\}} \langle v^{2}_{j}\rangle - \langle v_{j}\rangle^{2}}  \,. 
    \label{Equation:dispersion_velocity}
\end{equation}
The angled brackets indicate a mass-weighted average of the enclosed quantity. Measuring the dispersion using a cylindrical coordinate system removes the contribution of rotational motion that would otherwise be included in a randomly aligned Cartesian coordinate system. For this reason, as well as the fact that we use the three-dimensional information of the stellar particles, the `one-dimensional' velocity dispersion does not correspond to an observational line-of-sight velocity dispersion.

\subsection{Iterative versus non-iterative inertia tensors}\label{Subsection:IterativeInertiaTensors}

In this subsection we illustrate how the distribution of axis ratios for galaxies differs between using an iterative and non-iterative approach to compute the inertia tensor. The distribution of $c/a$ versus $b/a$ of central galaxies in {\midres} with $M_{*} \geq10^{9}\,\Msun$ ($\approx$ 500 bound stellar particles at this resolution) is shown in Fig.~\ref{Figure:inertia_tensor_joint}. The distributions shown in the left column use an iterative method, while those on the right column do not use iterations to compute the inertia tensors.

Since we expect disc morphologies to dominate for galaxies with stellar masses of $\approx 10^{10}\Msun$ \citep[e.g.][]{Clauwens.2018, Tacchella.2019}, we overlay contours that indicate the regions that enclose 75, 50 and 25 per cent of the galaxies in a sample with $10^{10}\leq M_{*}\,[\Msun] \leq 5 \times 10^{10}$ (20597 galaxies total). To facilitate the comparison between different choices, we divide the inertia tensor parameter space into three `morphological' classes following \citet{vanderWel.2014}. We use these regions, which are indicated by the dotted lines in Fig.~\ref{Figure:inertia_tensor_joint}, to measure the fraction of disc, triaxial and prolate galaxies ($f_{\rm disc}$, $f_{\rm triaxial}$, $f_{\rm prolate}$, respectively) in the $M_{*} \sim10^{10.3}\,\Msun$ sample.

Focusing on the first row, which uses an initial aperture equal to the half-stellar-mass radius, it is clear that the measured distributions differ between the iterative and non-iterative methods. There are nearly an order of magnitude more `triaxial' galaxies when using the non-iterative method compared to the iterative approach. This is because the non-iterative method yields overall larger values for both $b/a$ and $c/a$, i.e. more spherically symmetric systems. For example, the lowest values of $b/a$ are $\approx0.2$ and $\approx0.4$ in the iterative and non-iterative methods, respectively. Similarly, the lowest value of $c/a$ reaches $\approx0.1$ using iterative inertia tensors and $\approx0.3$ when no iterations are done. Consequently, the fraction of disc galaxies is much lower when using the non-iterative method ($f_{\rm disc} = 0.44$) relative to the iterative version ($f_{\rm disc} = 0.69$), simply because of the symmetry enforced by a spherical aperture.

The distributions of axis ratios inferred from the two approaches become increasingly similar as larger apertures are used. The last row of Fig.~\ref{Figure:inertia_tensor_joint} uses all bound stellar particles to define the aperture size, at which point the shapes of the distribution are the most similar between the two methods, particularly along the $c/a$ direction. Note however that there are still more discs in the iterative ($f_{\rm disc} = 0.59$) than in the non-iterative ($f_{\rm disc} = 0.42$) approach. As we argue in the following subsection, an aperture that is too large relative to the galaxy size makes the inertia tensor sensitive to the diffuse envelope of stars that are kinematically hotter and less disc-like. Hence, the bias induced by using a spherically symmetric aperture lessens if the input particle distribution becomes truly more spherically symmetric.

\subsection{The choice of aperture size}\label{Appendix:ApertureSize}

\begin{figure}
    \centering
    \includegraphics[width=\columnwidth,keepaspectratio]{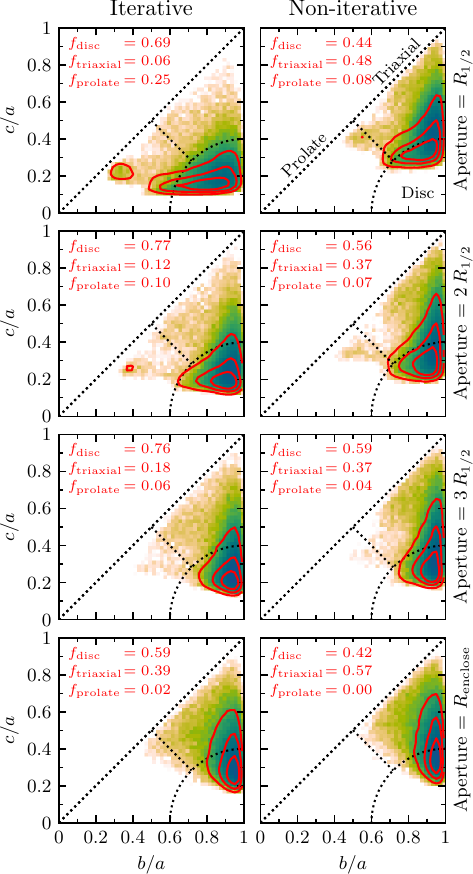}
    \caption{Distribution of axis ratios for {\tt L200m6} central galaxies with $M_{*}\geq10^{9}\,\Msun$. The dotted lines delineate regions of approximately similar morphological types \citep[][]{vanderWel.2014}, as indicated in the top right panel. The panels in the left column use the iterative method to measure the inertia tensor, whereas those on the right employ the non-iterative approach. The initial aperture used to select stellar particles in the first, second and third rows is one, two and three times the half-stellar-mass radius of the galaxy respectively, as indicated by the right-hand-side labels. The bottom row uses the smallest aperture that encloses all bound stellar particles. The overlaid contours enclose 75, 50 and 25 per cent of galaxies with $10^{10}\leq M_{*}[\Msun]\leq5\times10^{10}$, after smoothing the number count with a Gaussian kernel with a constant covariance of 0.15. The disc, triaxial and prolate fractions of galaxies within this stellar mass range are indicated in the top left corner of each panel. The iterative method results in a larger fraction of disc galaxies than the non-iterative approach. Apertures that are too large make the inertia tensors sensitive to structures that are external to galaxies, whereas using too small apertures underestimates the contribution of the disc, misrepresenting the shape of the galaxy itself.}
    \label{Figure:inertia_tensor_joint}
\end{figure}

Measuring any of the morphology metrics we define for this study requires an input particle distribution. As such, several choices can be made when it comes to selecting which particles are taken into account. We limit our selection to bound stellar particles, as it will prevent contamination from nearby galaxies, allowing us to focus on the actual intrinsic morphology of the galaxy under consideration. Another choice is the size of the spatial aperture. Using an aperture that is small compared to the overall extent of the galaxy artificially increases the relative importance of inner structures, like stellar bars and bulges. Conversely, using a very large aperture picks up stars that belong to the stellar halo. Between both extremes lies an aperture choice that encompasses most of any potentially present disc.

To decide which aperture size works best, we explore the effect of changing the initial aperture on the distribution of axis ratios derived from 3D mass-weighted iterative inertia tensors. We explore the effect of changing the aperture using a spatial metric instead of a kinematic metric for two reasons. First, the value of a kinematic metric could change as a function of aperture simply because the reference angular momentum vector becomes misaligned compared to the disc. For example, two co-spatial discs with opposite angular momenta may be spatially thin but the kinematic metrics may measure a large non-rotating component depending on which component dominates within the aperture (see \S\ref{Appendix:LuminosityWeighting}). Second, it is more straightforward to identify a sensible size aperture when using a spatial metric. Using an aperture that is too large or too small will result in many triaxial or prolate galaxies, whereas one expects disc galaxies to have roughly $a \approx b \gg c$. Hence, the aperture most suitable for our purposes is the one that is sufficiently large to encompass most of the disc and is insensitive to internal disc structures, but not so large that it becomes sensitive to accreted material.

As for the iterative versus non-iterative inertia tensor comparison, we only include central galaxies from {\midres} with at least $10^{9}\,\Msun$ in stellar mass. We use four different initial apertures. The first three correspond to one, two and three times the half-stellar-mass radius. The fourth aperture is the smallest radius that encloses all bound stellar particles, $R_{\rm enclose}$. Note that, since we compute inertia tensors iteratively, some particles can be further away from the initial spherical cut, as the sphere is reshaped into an ellipsoid whose volume remains fixed across iterations. The resulting axis ratios for each aperture choice are shown across the different rows of Fig.~\ref{Figure:inertia_tensor_joint}. 

Starting with the smallest aperture, and only considering the iterative inertia tensor from now on, all regions of the parameter space include a substantial fraction of galaxies. The large majority of galaxies in the triaxial region of the parameter space are galaxies with stellar masses below $10^{10}\,\Msun$ or above $10^{11}\,\Msun$. Focusing on the contours, we see that the mass-selected sample ($10^{10}\leq M_{*}\,[\Msun] \leq 5\times10^{10}$) preferentially has a low $c/a$, but has nonetheless a very broad distribution in $b/a$. In fact, there appear to be two distinct populations of galaxies in the selected stellar mass range, regardless of whether iterative or non-iterative inertia tensors are used. One population has disc-like shapes and one has prolate-like shapes. There is a connecting bridge of galaxies between the peaks in the axis ratio distribution, but they are less numerous.

\begin{figure}
    \centering
    \includegraphics{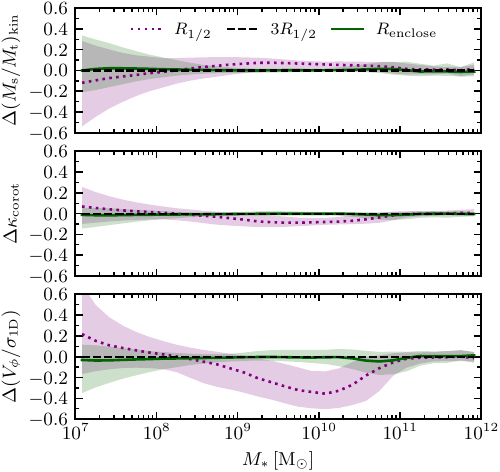}
    \caption{The effect of varying the aperture size on kinematic morphology metrics as a function of stellar mass. The lines show the median shift of a given morphology metric relative to our fiducial aperture of $3R_{1/2}$, and the shaded regions indicate the 16th to 84th percentiles of the distributions. We only use central galaxies from the {\midres} simulation. Using an aperture that is too small underestimates the kinematic importance of discs.}
    \label{Figure:median_shift_metrics_aperture_effect}
\end{figure}

Imaging the galaxies associated to the concentration of prolate galaxies in the top left panel of Fig.~\ref{Figure:inertia_tensor_joint} reveals that they are mostly strongly barred disc galaxies. Hence, their prolate classification originates from the dominant contribution of the bar, rather than the global shape of the galaxy. Thus, using the half-stellar-mass radius is suboptimal because it is clearly sensitive to the inner structure of the galaxy, rather than its overall stellar mass distribution. Increasing the aperture to twice the half-stellar-mass radius mitigates this problem, as the fraction of prolate galaxies decreases. However, the population of strongly barred galaxies still remains as an overdensity in the distribution of galaxies in the second row of the first column of Fig.~\ref{Figure:inertia_tensor_joint}. It is only when using three times the half-stellar-mass radius (third row of the left column of Fig.~\ref{Figure:inertia_tensor_joint}) that the feature largely disappears.

The fraction of disc galaxies in the $10^{10}\leq M_{*}\;[\Msun]\leq 5\times10^{10}$ regime increases as we increase the aperture, from $R_{\rm 1/2}$ ($f_{\rm disc} = 0.69 $) to $2R_{\rm 1/2}$  ($f_{\rm disc} = 0.77$). The disc fraction decreases very slightly to 0.76 when using $3R_{\rm 1/2}$. The changes in the disc fraction for these three apertures are minor compared to when the enclose radius is used, in which case the disc fraction decreases to 0.59. The notable decrease in the disc fraction is due to a large fraction of these mass-selected galaxies being classified as triaxial. We attribute this to the fact that we are now sensitive to the stellar halo of these galaxies, which is more triaxial. This is partly due to our definition of the inertia tensor, $I \propto r^{2}$. Distant particles therefore contribute more to the total inertia tensor than nearby particles (see \citealt[][]{Zemp.2011} for a comparison of the effect of different particle-weighting schemes). Given all of the above, we choose $3R_{\rm 1/2}$ as our fiducial aperture because it is the smallest value that is relatively insensitive to both the central and extended structure of the galaxy.

To conclude, we show in Fig.~\ref{Figure:median_shift_metrics_aperture_effect} how much the kinematic metrics vary as we change the aperture from our fiducial choice of $3R_{1/2}$ to the smallest and largest apertures considered in this analysis. Using all bound particles ($R_{\rm enclose}$; solid line) does not result in a significant systematic shift in the median values, although for some stellar masses the scatter becomes asymmetric and preferentially moves galaxies towards the dispersion-dominated regime. Using $R_{1/2}$ (dotted line) has larger effects, particularly in the stellar mass range between $2\times 10^{8}\Msun$ and $10^{11}\Msun$, where galaxies become more dispersion-supported for this aperture. Below $2\times 10^{8}\Msun$, a smaller aperture actually increases the degree of rotational support, and the scatter becomes large. This is likely caused by having an insufficient number of stellar particles, because the half-stellar-mass radius in this stellar mass range encloses less than a few tens of particles. As we show in \S\ref{convergence_sampling_size}, having an insufficient number of sampling particles biases the morphology metrics towards more disc-like values. Overall, the metric that is the most sensitive to the chosen aperture is $(V_{\phi}/\sigma_{\rm 1D})$. All other morphology metrics do not vary by much as the size of the aperture changes. 

\subsection{The effect of luminosity weighting}\label{Appendix:LuminosityWeighting}

\begin{figure}
    \centering
    \includegraphics{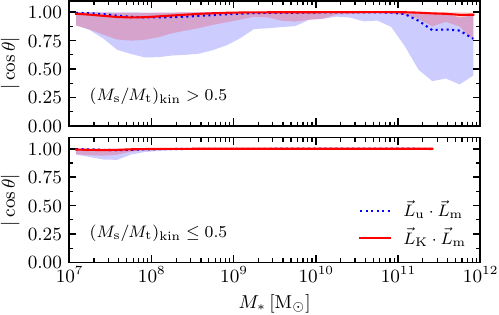}
    \caption{Misalignment between the mass- and luminosity-weighted stellar angular momentum within $3R_{1/2}$ as a function of stellar mass. The median misalignment when using the u and K GAMA photometric bands are shown using the dotted and solid lines, respectively. The shaded regions indicate the 16th to 84th percentiles of the distributions. Galaxies are split into spheroid- (top panel) and disc-dominated (bottom panel) samples according to their mass-weighted spheroid-to-total mass ratios. We only use central galaxies from the {\midres} simulation. The mass- and luminosity-weighted angular momenta are well aligned, particularly for disc-dominated galaxies.}
    \label{Figure:angular_momentum_misalignment}
\end{figure}

In this work we mainly use mass-weighted quantities. However, the spatial distribution and kinematics of stars can differ in a manner that correlates with their ages. Younger and bluer stars are generally expected to form within rotationally-supported gas discs, whereas old and redder stars may be kinematically hotter as a result of mergers and interactions over the lifetime of their host galaxy. The use of different photometric bands may thus systematically change the apparent morphology of the galaxy, which may differ from the one inferred from the stellar mass distribution. 

To quantify the effect of luminosity-weighting on our chosen morphology metrics, we explore how their values differ compared to their mass-weighted definitions when weighting each stellar particle by its fractional contribution towards the total luminosity in a given band:
\begin{equation}
    w_{i} = L_{i} / \sum\limits^{N}_{j}L_{j} \, .
\end{equation}
The mass term $m_{i}$ appearing in equations \ref{Equation:inertia_tensor}, \ref{equation:kappa_corot}, \ref{Equation:spheroid_to_total} and \ref{Equation:rotational_velocity} is subsequently replaced by $w_{i}$. We also use luminosity-weighted averages for Eq. \ref{Equation:dispersion_velocity}. Note that we opt to use three times the half-stellar-mass radius as our aperture, rather than three times the half-stellar-light radius. This choice is made to ensure that any changes we observe are only driven by the luminosity weighting of particles, rather than by an implicit change in the aperture size, as the half-stellar-light radius will differ across bands and from the half-stellar-mass radius.

For the purposes of this comparison, we only consider the reddest (K) and bluest (u) GAMA \citep[][]{Driver.2009} photometric bands. The K band is less biased towards younger stars, whereas the u band is significantly biased towards young stellar populations. Hence, using other GAMA bands will result in differences that lie in between the results obtained when using the K or u bands. The intrinsic luminosity of each stellar particle in these two bands is obtained using the method described in \citet[][]{Trayford.2015}, which does not take into account the effect of dust. In short, the spectral energy density (SED) of stellar particles depends on their metallicity and age, which are obtained using interpolated tables based on the {\tt GALAXEV} stellar evolution model \citep[][]{Bruzual.2003}. The SED is then convolved with the corresponding photometric broadband filter. Since we do not include dust when computing GAMA luminosities, the effect of u-band luminosity-weighting is likely overestimated relative to observations, as young stellar populations are often enshrouded in dusty birth clouds. Since dust lanes preferentially attenuate emission from the galaxy mid-plane, their presence can have important implications for the measured vertical thickness of galaxies (i.e. the difference between observed and intrinsic vertical scale heights is larger for shorter wavelengths; e.g. \citealt{Bizyaev.2009, Yu.2026}).

Before exploring how the morphology metrics change between luminosity- and mass-weighted definitions, we check whether the alignment of the luminosity-weighted angular momentum differs from its mass-weighted counterpart. Since the angular momentum vector is used as a reference direction from which kinematic metrics are computed, a change in its alignment would lead to a change in the value of a given morphology metric.

We show in Fig.~\ref{Figure:angular_momentum_misalignment} the average misalignment angle between the mass- and luminosity-weighted angular momentum as a function of stellar mass for spheroid- (top panel) and disc-dominated (bottom panel) central galaxies in the {\midres} simulation. We constrain the values of $\cos \theta$ to be positive by taking its absolute value, but we note that a small fraction ($\approx 5 \times 10^{-5}$, depending on the chosen band) of galaxies have $\cos \theta \approx -1$. An antiparallel angular momentum vector could indicate the existence of counter-rotating stellar discs, whose origin, properties and incidence in COLIBRE galaxies is beyond the scope of this study. We see that disc-dominated galaxies have well-defined angular momentum directions, as only galaxies with $M_{*} \leq 10^{8}\,\Msun$ ($\approx 100$ stellar particles) deviate from $\cos \theta=1$. On the other hand, spheroid-dominated galaxies exhibit a larger scatter, which is because they have low angular momentum to begin with. Hence, a small change in the angular momentum can change the net direction of $\vec{L}_{*}$ substantially. 

\begin{figure}
    \centering
    \includegraphics{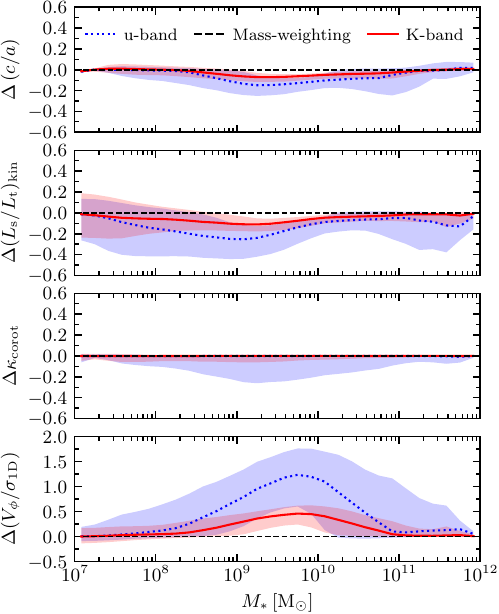}
    \caption{Comparison of mass- and luminosity-weighted morphology metrics measured within $3R_{1/2}$. The lines indicate the median shift of a given morphology metric relative to its mass-weighted definition at a fixed stellar mass, when using the u (dotted) and K (solid) GAMA photometric bands. The shaded regions indicate the 16th to 84th percentiles of the distributions.  We only use central galaxies from the {\midres} simulation. Using luminosity-weighted morphology metrics results in flatter and more disc-dominated galaxies than using mass-weighted morphology metrics.}
    \label{spheroid_to_total_mass_vs_luminosity_weighting}
    \label{Figure:luminosity_weighting}
\end{figure}

Because of the relatively small amount of misalignment between the mass- and luminosity-weighted stellar angular momenta, measurements reliant on comparing the amount of counter- and co-rotating mass will not differ significantly. Hence, the spheroid-to-total mass ratio measured using luminosity-weighted angular momenta will remain unchanged, which we have verified. Nonetheless, this does not preclude differences in the spheroid-to-total luminosity ratio, $(L_{\rm s}/L_{\rm t})_{\rm kin}$. 

\begin{figure*}
  \centering
  \includegraphics[width=\textwidth, keepaspectratio]{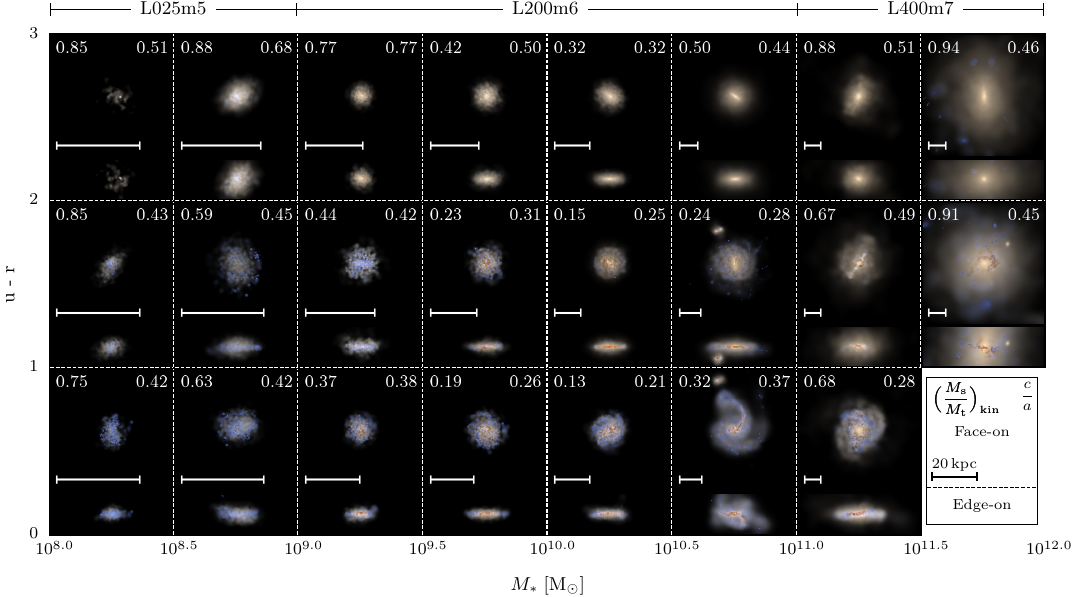}
  \caption{HST-like images of representative galaxies from the {\highres}, {\midres} and {\lowres} simulations, which are used for different stellar mass ranges, as indicated at the top of the figure.  The images use all particles in the region regardless of whether they are bound or not to the galaxy of interest, meaning that nearby companion galaxies may also be imaged. Since we compute morphology metrics using only bound stellar particles, companion galaxies do not affect the values measured for the main galaxy in the image. The stellar mass of galaxies increases from left to right and their intrinsic colours become bluer from top to bottom. Each galaxy was chosen based on how close its minor-to-major axis ratio and spheroid-to-total mass ratio are to the median value of galaxies within the same mass and colour bin. We show a face- and edge-on view, determined from the stellar angular momentum within $3R_{1/2}$. There are no galaxies bluer than (u-r)~$ \leq 1$ above $M_{*} = 10^{11.5}\,\Msun$ in {\lowres}.The layout and elements of each panel are illustrated in the bottom right legend.}  
    \label{Figure:example_images}
\end{figure*}

We show in Fig.~\ref{Figure:luminosity_weighting} the shift in the values of morphology metrics for our chosen luminosity bands, compared to using mass-weighted quantities. The median $\kappa_{\rm corot}$ values remain nearly the same as for the mass-weighted approach, although the scatter is asymmetric and leads to slightly more spheroid-dominated galaxies. The other metrics are more sensitive to the effect of luminosity-weighting, leading to flatter galaxies with more rotational support relative to a mass-weighted approach (particularly for the u-band). The metric showing the largest change is $V_{\phi}/\sigma_{\rm 1D}$, which is primarily driven by an increase in the rotational velocity, $V_{\phi}$, rather than a decrease in the velocity dispersion, $\sigma_{\rm 1D}$.

In short, the results that we present in this paper likely represent a lower limit on how rotationally-dominated or thin COLIBRE galaxies would be if observed. Using luminosity-weighting instead of mass-weighting decreases the stellar mass above which disc galaxies dominate. Although not shown here, the mass threshold of having 50 per cent disc galaxies is crossed at $\approx2\times10^{9}\,\Msun$ when using a $(M_{\rm s}/M_{\rm t})_{\rm kin} \leq 0.5$ cut. In contrast, using $(L_{\rm s}/L_{\rm t})_{\rm kin} \leq 0.5$ for the K band (u band) results in a threshold of $M_{*}\approx7\times10^{8}\,\Msun$ ($M_{*}\approx3\times10^{8}\,\Msun$). Although luminosity-weighting modifies $V_{\phi}/\sigma_{\rm 1D}$ by the largest amount, the change primarily occurs in the range where disc galaxies already dominate. The threshold stellar mass above which 50 per cent of galaxies have $V_{\phi}/\sigma_{\rm 1D} \geq 1$ shifts from $M_{*}\approx 3\times10^{9}\,\Msun$ when weighting by mass to $M_{*}\approx 10^{9}\,\Msun$ ($M_{*}\approx 5\times10^{8}\,\Msun$) when weighting by the K band (u band) luminosity.

\section{Results}\label{Section:results}

We begin this section by visualising, in \S\ref{section:example_images}, several example galaxies taken from the {\lowres}, {\midres} and {\highres} simulations. They were selected to illustrate the diverse range of morphologies that exist in the {\tt COLIBRE} simulations. We then quantify, in \S\ref{Section:ConvergeMorphology} and \S\ref{Section:MorphologicalCorrelations}, how well-converged and correlated our chosen morphology indicators are. Following this, in \S\ref{Section:CorrelationMorphologyStellarMass}, we relate the morphology of galaxies to their stellar masses and environment. In \S\ref{Section:CorrelationMorphologyStellarMassColour} we look at the relation between morphology and galaxy intrinsic colour. Lastly, in \S\ref{Section:CorrelationMorphologyHaloProperties} and \S\ref{Section:CorrelationMorphologyOtherGalaxyProperties}, we quantify the correlation strength between galaxy morphology and various halo and galaxy properties.

\subsection{Visual impressions of galaxy morphologies}\label{section:example_images}

We show several images of galaxies from the {\highres}, {\midres} and {\lowres} simulations in Fig.~\ref{Figure:example_images}, which are coloured using the \textit{Hubble Space Telescope} (HST) filters of \citet{Lupton.2004} for the ACS/WFC F475W (blue), F625W (green) and F775W (red) bands. The galaxies are arranged according to their stellar mass, which increases from left to right, and their intrinsic (dust-free) colour, which becomes bluer from top to bottom. We use different resolution simulations in different stellar mass ranges, so that the lowest mass galaxies shown for any given resolution still contain more than $10^{3}$ stellar particles. The {\highres} simulation is used for galaxies with $10^{8} \leq M_{*} [\Msun] < 10^{9}$, {\midres} from $10^{9} \leq M_{*} [\Msun] < 10^{11}$ and {\lowres} from $10^{11} \leq M_{*} [\Msun]\leq 10^{12}$. The galaxies were chosen because they have $(M_{\rm s}/ M_{\rm t})_{\rm kin}$ and $c/a$ values that are closest to the median values of galaxies of similar mass and colour.

The images were generated using the procedure described in Hu\v{s}ko et al. (in prep). Briefly, a 3D Cartesian grid is overlaid around the centre of each galaxy, and the dust and stellar luminosity local to each cell is computed by SPH-interpolating the relevant fields from the gas and stellar particles. The line-of-sight dust surface mass density is used with a 1D radiative transfer model that takes into account dust chemical composition and grain sizes. The model then calculates the amount of light attenuation, caused by absorption and scattering (the latter assuming extinction only), as function of wavelength. Although this method does not account for light scattered into the line-of-sight, its contribution is minor.

The variety of galaxy morphologies in COLIBRE is evident from Fig.~\ref{Figure:example_images}. Galaxies with $M_{*} \leq 10^{9}\,\Msun$ are dominated by a spheroid component, although their shapes can become flatter if they are actively star forming. There is a relatively small amount of dust obscuration in this stellar mass range, so there is a good correspondence between intrinsic and observed colours. The leftmost two columns also show that very different spheroid-to-total mass ratios can correspond to similar $c/a$ values. We will investigate the correlation and scatter between shape and kinematic morphology metrics more closely in \S\ref{Section:MorphologicalCorrelations}.

As we progress towards the massive dwarf galaxy regime ($10^{9} \leq M_{*} \, [\Msun]\leq 10^{10}$), we observe a systematic increase in the importance of stellar discs, which propagates to flatter stellar mass distributions. The diversity of galaxy morphologies at a fixed stellar mass also increases compared to lower stellar mass dwarf galaxies. The reddest galaxies in this stellar mass range (u-r$\geq 2$) are typically spheroid-dominated and thick, whereas the bluest galaxies (u-r$\leq 1$) are disc-dominated and thin. 

Most galaxies are disc-dominated between $10^{10} \leq M_{*} \, [\Msun]\leq 10^{11}$, which is also the mass range where galaxies are the thinnest. The disc component is important regardless of the colour of the galaxy, as we also see red disc galaxies, of which some contain visibly strong stellar bars. Notably, the bluest galaxies (u-r$\leq 1$) with $10^{10.5} \leq M_{*} \, [\Msun]\leq 10^{11}$ often exhibit signs of ongoing or past interactions with neighbouring galaxies.

The most massive galaxies in our sample, $M_{*}\geq 10^{11}\,\Msun$, are spheroid-dominated. There are nonetheless some trends with colour that mimic what was found for the blue galaxies with $10^{10.5} \leq M_{*} \, [\Msun]\leq 10^{11}$. Some of the bluest galaxies in this stellar mass range appear to have recently been disturbed but remain quite flat, like the example shown in the last row of the second-to-last column. Other galaxies, like the one shown in the bottom panel of the rightmost column, also contain gas discs that are perpendicular to the major axis of the stellar mass distribution, suggesting that the gas was recently accreted in a misaligned fashion and creating a polar-ring-like galaxy. Although these gas discs usually appear prominent, and can contain some amount of star formation, the stellar mass content of the disc is minor relative to that of the entire galaxy.

\subsection{Numerical convergence of morphology}\label{Section:ConvergeMorphology}

Quantifying the numerical convergence of the morphology indicators is an important step in determining which galaxies have robustly quantifiable morphologies. An important subtlety is that morphology measurements of simulated galaxies can generally be influenced by three separate but related numerical effects.

The first aspect is the number of particles that sample the underlying stellar phase-space distribution. Reducing the number of tracers increases the importance of shot noise, which can eventually bias the measured morphologies. This sampling effect has been investigated in the context of how satellite subhaloes trace the shapes of their host dark matter haloes, e.g. \citet[][]{Herle.2025}, but since the morphologies of galaxies are more varied than those of dark matter haloes, we explicitly verify this effect in \S\ref{convergence_sampling_size} for central galaxies in COLIBRE. This exercise allows us to identify the particle number below which morphology measurements become biased.

The second aspect is the fact that changing the numerical resolution of the simulation changes the birth properties of stars, and therefore the properties of galaxies \citep[][]{Ludlow.2020}. An increase in the resolution generally enables the formation of denser gas, which shifts star formation towards higher densities and lower temperatures (e.g. fig. 10 of \citealt{Schaye.2025}). Furthermore, as a consequence of the changes in the physical scales that are resolved, changing the resolution often requires some adjustments to the parameters governing the subgrid physics in order to maintain agreement with the observables used for the calibration. The attempt to match the predictions for present-day galaxy masses and sizes across resolutions to observational data is not guaranteed to be successful, and even if it is successful, it does not guarantee that other observables are converged with the simulation resolution (e.g. galaxy morphologies). Hence, we also check in \S\ref{convergence_simulation_resolution} how the morphologies of central galaxies vary across the {\tt m7}, {\tt m6} and {\tt m5} models.

The third effect is driven by the spurious heating of the stellar component by gravitational interactions with the surrounding DM. This process becomes important when the number of particles is small enough for the relaxation times to be shorter than the Hubble time. Affected galaxies thus experience unphysical kinematic heating throughout their evolution, leading to untrustworthy morphologies for a given galaxy formation model. Contrary to the first two effects discussed above, it is difficult to quantify how important this effect is in COLIBRE. Morphological differences between different resolution simulations could differ because the lower resolution simulation is more strongly affected by numerical artefacts than the higher resolution one, or because the galaxy morphologies are intrinsically different because of a different subgrid physics calibration.

\subsubsection{Convergence with particle sample number}\label{convergence_sampling_size}

We test the influence of sampling noise on the measured morphology parameters of galaxies as follows. First, we compute the metrics defined in \S\ref{Section:MorphologicalCalculations} for all galaxies in the {\midres} simulation box. Second, we re-compute those same metrics after randomly selecting one eighth of the bound stellar particles for each galaxy. When subsampling, we scale stellar masses of the selected particles to ensure mass conservation. This approach effectively produces an `{\tt m7}' representation of an `{\tt m6}' galaxy, whilst retaining the `{\tt m6}' stellar birth properties. Hence, this test serves to identify systematic differences that are caused solely by particle sampling noise.

\begin{figure}
    \centering
    \includegraphics{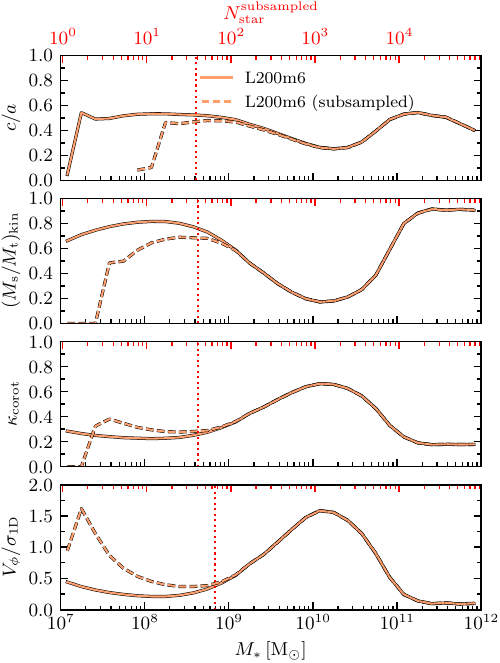}
    \caption{The effect of the number of stellar particles ($N_{\rm star}$) used to sample the underlying spatial and velocity distributions on the measured morphology of all galaxies in the {\midres} simulation. The solid lines show the median value of each morphology indicator as a function of the stellar mass of the galaxy. The dashed lines show how the median values change when randomly subsampling the stellar particles by a factor of eight, effectively mimicking an {\tt m7} resolution simulation with the same stellar birth properties as the `{\tt m6}' simulation. The stellar mass below which the median values first deviate from each other by more than 10 per cent is indicated by the vertical dotted lines. The correspondence between stellar mass and the average number of stellar particles at `{\tt m7}' resolution ($N^{\rm subsampled}_{\rm star}$) is indicated along the top x-axis. Poor sampling ($N_{\rm star} \lesssim 50$) leads to an overestimate of the importance of the disc within galaxies.}
\label{Figure:sampling_particle_convergence}
\end{figure}

We show in Fig.~\ref{Figure:sampling_particle_convergence} the stellar mass dependence of our four different morphology indicators and how they differ between the cases when all the bound stellar particles of galaxies are used (solid lines) and when we subsample the stars (dashed lines). Galaxies with stellar masses above $10^{10}\,\Msun$ are well-sampled, as the subsampling does not shift the median values. As we progress towards lower stellar masses, the effect of subsampling becomes increasingly noticeable.

Overall, poorly sampled galaxies tend to be flatter and have larger rotational support than the true underlying population. This somewhat counterintuitive trend can be understood by considering the case of how one may generate a discrete representation of a spherically symmetric distribution. For a sufficiently large number of sampling points, the discrete representation will be close to spherical. However, if one samples with too few points, the distribution is likely to become elongated along a randomly oriented direction, reducing the spherical symmetry. Similarly, the sum of a small number of random angular momentum vectors likely yields a non-zero total angular momentum vector due to intrinsic Poisson variance. This vector defines an artificial plane of rotation, regardless of whether the particles are physically orbiting in this plane or not. Thus, morphology metrics for disc galaxies are less sensitive to sampling effects than for elliptical galaxies, a similar effect seen in triaxial vs spherical haloes \citep[][]{Dubinski.1991}. We repeated the subsampling test after separating the galaxy population according to their true $(M_{\rm s}/M_{\rm t})_{\rm kin}$ values, confirming this is indeed the case.

We highlight using a vertical dashed line in each panel of Fig.~\ref{Figure:sampling_particle_convergence} the stellar mass below which the median morphology indicator of the subsampled distribution differs by more than 10 per cent from the true value. All morphology metrics show approximately the same level of convergence, as they deviate from the true values below $M_{*} \approx (4 - 5)\times 10^{8}\,\Msun$. We can convert these stellar masses into a minimum required particle number by using the mean stellar particle mass of {\midres} and multiplying it by eight to account for our subsampling. In short, only $\approx 40 - 50$ stellar particles are required for converged medians, which is fewer than the number corresponding to the 10 per-cent-level difference estimated assuming Poisson statistics (100 particles). The required number of particles is somewhat larger ($\approx70$) for $V_{\phi}/\sigma_{\rm 1D}$, but this is because the smaller values of $V_{\phi}/\sigma_{\rm 1D}$ result in larger measured fractional differences, even if the absolute difference is smaller than for the other metrics. We have performed the same tests for {\lowres}, for which we degrade the resolution to `{\tt L400m8}', and find similar results as those presented here. 

Note that this discussion pertains only to the convergence of the median morphology values, as the values of individual galaxies before and after subsampling can differ more. The scatter induced by subsampling on an individual galaxy-level increases as the stellar mass decreases (there are fewer particles to begin with) and is lower for disc-dominated galaxies than for spheroid-dominated ones (it is less likely to find even more massive discs for disc-dominated galaxies). The symmetrised 80th - 20th percentile scatter is $\Delta(M_{\rm s}/M_{\rm t})_{\rm kin}\approx 0.06$ at $M_{*}\approx10^{9}\,\Msun$ in the {\midres} simulation.   

\subsubsection{Convergence with the intrinsic simulation resolution}\label{convergence_simulation_resolution}

Our last test for numerical convergence investigates the effect of the initial particle masses. Contrary to subsampling, changing the particle masses directly in the simulation can modify the manner in which galaxy formation proceeds, thus affecting the birth properties and distributions of star particles. Thus, even for stellar mass ranges that are well sampled at a given resolution, the values of morphology indicators may change between the {\tt m5}, {\tt m6}, and {\tt m7} models.

\begin{figure}
    \centering
    \includegraphics{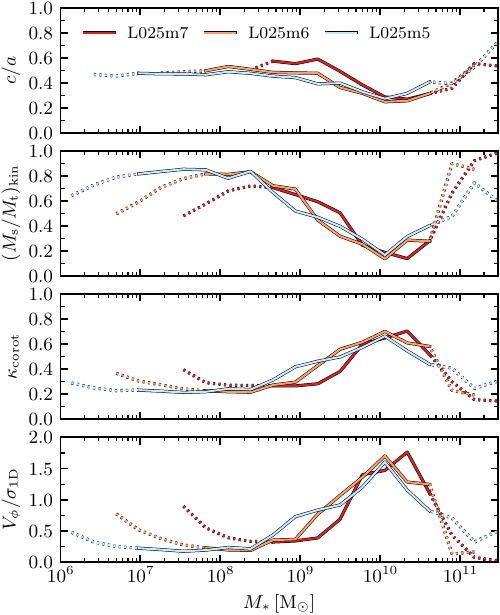}
    \caption{Convergence of the median values of four morphology indicators with the numerical resolution of the simulation. The tests are done using all galaxies from the largest available volume ($25^{3}$~Mpc$^{3}$) at $z=0$ common to the {\tt m5}, {\tt m6} and {\tt m7} models. The dotted lines at the high mass end indicate stellar mass bins that contain fewer than 10 galaxies. The stellar mass bins in which galaxies are poorly sampled ($\lesssim 50$ particles) are also indicated by dotted lines. Contrary to Fig.~\ref{Figure:sampling_particle_convergence}, this convergence test is sensitive to how the stellar birth properties change across different mass resolution models. The dependence of morphology on stellar mass is similar across all models, except when galaxies are poorly sampled. The poorly sampled regime appears as a spurious overestimate of disc importance, with the inflection point occurring at the stellar mass predicted from the analysis of Fig.~\ref{Figure:sampling_particle_convergence}}.
    \label{Figure:resolution_convergence}
\end{figure}

Fig.~\ref{Figure:resolution_convergence} shows the median value of the morphology metrics as a function of stellar mass at $z = 0$ for the $(25\,\mathrm{Mpc})^{3}$ volume. Note that the median curve for the minor-to-major axis ratio starts at a higher stellar mass than the other morphology metrics because we require a minimum of 20 stellar particles in the initial aperture for it to be computed, whereas no such constraint is present for the other metrics.

The stellar mass dependence of all morphology indicators is similar across the three resolutions. The most rotationally-dominated and flattest galaxies have stellar masses of $\approx 2\times10^{10}\,\Msun$. Above this stellar mass, there is a steep increase in the importance of dispersion support. Towards lower masses, galaxies also become increasingly spheroidal and dispersion-dominated until reaching a resolution-dependent mass threshold. Below this limit, the significance of ordered rotation and the prevalence of rotating discs artificially increase, which is likely driven by sampling noise. It resembles the dependence discussed in \S\ref{convergence_sampling_size} and the stellar mass at which it happens shifts towards lower values as the resolution of the simulation increases. 

Numerical resolution may also play a role in the morphology of galaxies if the number of particles is not high enough to prevent artificial heating of the stars by DM particles. Contrary to the effect of shot noise, this effect is more difficult to quantify because the intrinsic properties of galaxies may change between different resolutions. Hence, changes in the median morphology of galaxies at fixed stellar mass between resolutions can reflect spurious heating or different galaxy properties. For example, galaxies with $M_{*}\approx 10^{9}\,\Msun$ appear systematically thinner and have a smaller velocity dispersion as the resolution increases, but we note that it is difficult to quantify how robust small deviations between each simulation are given the relatively small cosmic volume of the $(25\,\mathrm{Mpc})^{3}$ box.

\begin{figure}
    \centering
    \includegraphics{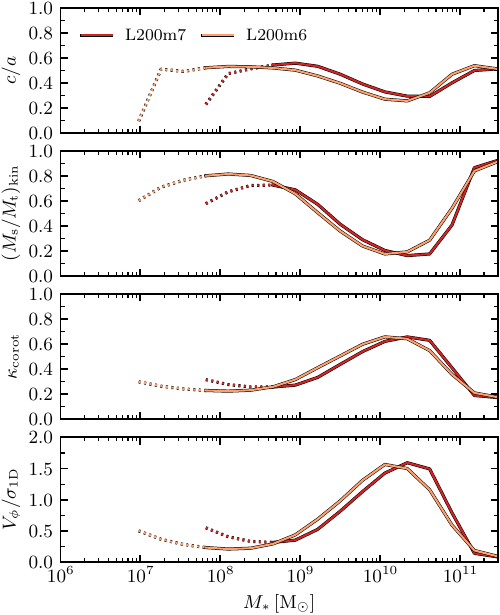}
    \caption{As Fig.~\ref{Figure:resolution_convergence}, but using the largest volume that is available at $z = 0$ for both the {\tt m7} and {\tt m6} resolutions.}
    \label{Figure:resolution_convergence_larger_box}
\end{figure}

For the convergence between {\tt m6} and {\tt m7}, we can use the $(200\,\mathrm{Mpc})^{3}$ volume to improve the statistics, which we show in Fig.~\ref{Figure:resolution_convergence_larger_box}. Doing so reveals that the difference in morphology at $M_{*}\approx 10^{9}\,\Msun$ is partly caused by a small but systematic shift towards higher stellar masses for the {\tt m7} morphology-to-stellar-mass relation relative to the one found in the {\tt m6} model. The most rotationally-dominated galaxies at {\tt m7} resolution have $M_{*}\approx2\times10^{10}\,\Msun$, whereas it is closer to $M_{*}\approx10^{10}\,\Msun$ for {\tt m6}. The shift also affects galaxies below and above this stellar mass range. Dwarf galaxies are slightly more rotationally-supported for {\tt m6} than {\tt m7}, and galaxies with $(2 - 10)\times10^{10}\,\Msun$ are more rotationally-supported at {\tt m7} resolution. The level of difference for central galaxies is still small, about $\Delta(M_{\rm s}/M_{\rm t})_{\rm kin} \approx 0.05$ for the most discrepant values. For satellite galaxies, the differences are larger and reach $\Delta(M_{\rm s}/M_{\rm t})_{\rm kin} \approx 0.1$. The morphology-stellar mass relation for {\tt L200m7} below 50 stellar particles is dominated by shot noise, and closely follows the relation obtained by subsampling {\tt L200m6} (dashed line of Fig.~\ref{Figure:sampling_particle_convergence}).

In short, the median morphology of galaxies as a function of their stellar mass appears to converge well. This relatively good level of agreement across different mass resolutions is not expected a priori, since morphology is not a calibration target and could hence have varied much more. This means that, at least based on the $z = 0$ morphology of galaxies, we can combine simulations of different resolutions to balance the need for sufficient particles to sample the stellar distribution of galaxies with the need for a sufficient number of galaxies.

\subsection{Correlations between morphology indicators}\label{Section:MorphologicalCorrelations}

\begin{figure}
    \centering
    \includegraphics[width=\columnwidth,keepaspectratio]{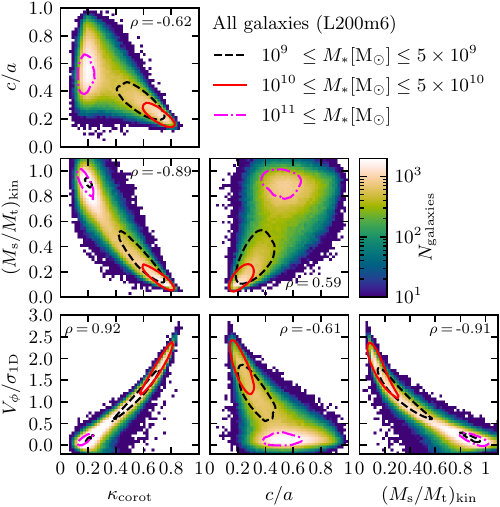}
    \caption{Correlations between four different indicators of morphology for all galaxies with $M_{*} \geq 5.4 \times 10^{7}\,\mathrm{M}_{\odot}$ in the {\midres} simulation (at least 50 stellar particles). The intensity of the map reflects the logarithm of the number of galaxies in a given pixel. The different contours enclose 50 per cent of the galaxy population within the three stellar mass bins indicated in the legend, which clearly differ in their morphology distributions. The contours are estimated using the galaxy number count for each mass-selected sample after smoothing it with a Gaussian kernel of constant covariance of 0.15. The Spearman rank correlation coefficient between the plotted metrics, $\rho$, is shown in each panel. All morphology metrics correlate strongly, particularly the kinematic ones.}
    \label{Figure:correlation_morphological_metrics}
\end{figure}

Having established a fiducial framework for computing morphology metrics, and having quantified their numerical convergence, we explore how correlated their values are. This is an interesting aspect to consider, since it not only shows how the spatial distribution of stars relates to their kinematics, but also tells us whether we can use a single morphology metric in lieu of using all four.

Fig.~\ref{Figure:correlation_morphological_metrics} shows the galaxy distributions between different mass-weighted morphology indicators for all galaxies in the {\midres} simulation with $M_{*} \geq 5.4 \times 10^{7}\,\Msun$. This mass threshold corresponds to galaxies containing at least 50 stellar particles at this resolution, satisfying the convergence requirements from \S\ref{convergence_sampling_size}. We overlay contours for three different stellar mass bins that indicate which regions of the parameter space enclose 50 per cent of three mass-selected subsamples: $10^{9}\leq M_{*}\,[\Msun] \leq 5\times10^{9}$, $10^{10}\leq M_{*}\,[\Msun] \leq 5\times10^{10}$ and $10^{11}\leq M_{*}\,[\Msun] \leq 5\times10^{11}$. The contours were found by minimising the area that encloses half of the corresponding galaxy population, after smoothing the number count distribution using a Gaussian kernel with a constant covariance equal to 0.15.

Focusing on the distribution as a whole, all metrics are very well correlated, as found in other simulations \citep[][]{Thob.2019}. The strongest correlations occur between kinematic quantities. Using the Spearman rank correlation coefficient ($\rho$) to quantify the correlation strength, we see that $\kappa_{\rm corot}$ ($\rho = -0.62$), $V_{\phi}/\sigma_{\rm 1D}$ ($\rho =-0.61$) and $(M_{\rm s}/M_{\rm t})_{\rm kin}$ ($\rho =0.59$) have similar correlation strengths with $c/a$. For all correlations between kinematic indicators, we find $|\rho| \geq 0.89$. We note that the correlation strength between pairs of metrics increases by $|\Delta\rho|=0.02$ when only considering central galaxies. 

The strength of the correlation between morphology metrics is sensitive to the stellar mass range under consideration. The correlation between kinematic metrics remains high ($|\rho| \geq 0.9$) between $M_{*} \approx 10^{9}\,\Msun$ and $M_{*} \approx10^{11}\,\Msun$. For less massive galaxies, the correlation strength steadily decreases, reaching $|\rho| \approx 0.5$ at $M_{*} = 7\times10^{7} \Msun$. A similar trend exists for the correlation strength between $c/a$ and the kinematic metrics. They correlate the strongest ($\rho \approx 0.7$) between $M_{*} \approx10^{9}\,\Msun$ and $M_{*} \approx 4\times10^{10}\,\Msun$ but weaken at higher and lower stellar masses. Out of the three kinematic metrics, $(M_{\rm s}/M_{\rm t})_{\rm kin}$ exhibits the weakest (albeit still strong) correlation with $c/a$. This is partly caused by an intrinsic limitation of this spheroid mass definition, as we explain below.

As mentioned briefly in the discussion of Fig.~\ref{Figure:angular_momentum_misalignment}, there are galaxies with counter-rotating stellar discs. For such galaxies, our definition of spheroid mass occasionally misrepresents the spatial morphology. For example, galaxies may have a seemingly unphysical value of $(M_{\rm s}/M_{\rm t})_{\rm kin}>1$. This occurs when the least massive disc out of two co-spatial counter-rotating discs has a larger angular momentum than the more massive one. This can happen because angular momentum is a combination of mass and distance to the centre of the galaxy, so if the spatial distribution of one disc is more extended than the other one, it can define the reference angular momentum vector of the whole galaxy, regardless of which of the two discs is more massive. Thus, in such cases, the mass of counter-rotating stars will be more than half of the total mass of the galaxy. Note that something similar can happen with $V_{\phi}$, which is why the lowest values of $V_{\phi}/\sigma_{\rm 1D}$ in Fig.~\ref{Figure:correlation_morphological_metrics} are negative. Since the definition of $\kappa_{\rm corot}$ is based on kinetic energy, and hence involves strictly positive values, it does not suffer from this issue.

Leaving aside the problems that any particular definition may have, part of the scatter between morphology indicators is determined by differences in the definition of parameters. Despite both being kinematic-derived parameters, there is for example a non-negligible amount of scatter in $(M_{\rm s}/M_{\rm t})_{\rm kin}$ at a fixed value of $\kappa_{\rm corot}$, even when excluding the extreme cases of counter-rotating galaxies. Part of the scatter is driven by galaxy properties that we have not included in our set of morphology parameters. For example, by selecting galaxies with similar levels of rotational support in a narrow range of stellar mass, we find that the value of $c/a$ correlates with the size of the galaxy with a strength of $\rho \approx 0.3$.

Regardless of its origin, the presence of scatter implies that using a fixed cut of, say, $\kappa_{\rm corot} \geq 0.4$ (as for example done in \citealt{Correa.2017}), can select galaxies with $(M_{\rm s}/M_{\rm t})_{\rm kin} \geq 0.5$. Hence, we caution against a selection based on a hard cut in the value of any given parameter. There is no \textit{a priori} value motivated by observations (observers do not measure this quantity) and there is some scatter relative to visual morphology \citep[e.g.][]{Pfeffer.2023}. Additionally, a given cut may not work well across different simulations, since the intrinsic distributions may differ depending on the galaxy formation model. With these caveats in mind, we would still like to identify a default morphology metric for our study. What metric out of our set minimises the sensitivity to operational choices, whilst also maximising the selection of galaxies that look like it and that have orbits consistent with discs?

Since we want to identify disc galaxies with disc-like orbits, we use a kinematic indicator, which, compared to shapes, correlates more strongly with the star formation rates and stellar ages of galaxies, both in simulations \citep[][]{Thob.2019} and observations \citep[][]{vandeSande.2018, Wang.2025}. Nonetheless, we still want to choose a metric that reflects the shape of the galaxy. The spatial distribution of galaxies as measured by $c/a$ correlates the strongest with $V_{\phi}/\sigma_{\rm 1D}$. One may therefore be inclined to choose $V_{\phi}/\sigma_{\rm 1D}$ as the fiducial metric. However, recall from the bottom panel of Fig.~\ref{Figure:median_shift_metrics_aperture_effect} and Fig.~\ref{Figure:luminosity_weighting} that this metric is the most sensitive to the choice of aperture and luminosity-weighting. Given its dependence on operational choices, it is undesirable to use it as the fiducial metric. The second most strongly correlated metric is $\kappa_{\rm corot}$. Contrary to $V_{\phi}/\sigma_{\rm 1D}$, it is insensitive to luminosity-weighting and the choice of aperture. Nonetheless, the fraction of kinetic energy in ordered co-rotation is not readily available from observations. Hence, the last option is the spheroid-to-total stellar mass ratio. Despite its shortcomings when it comes to the small fraction of counter-rotating discs, we prefer using this kinematic metric as our default whenever we want to limit the morphology metrics to just one.

\begin{figure}
    \centering
    \includegraphics{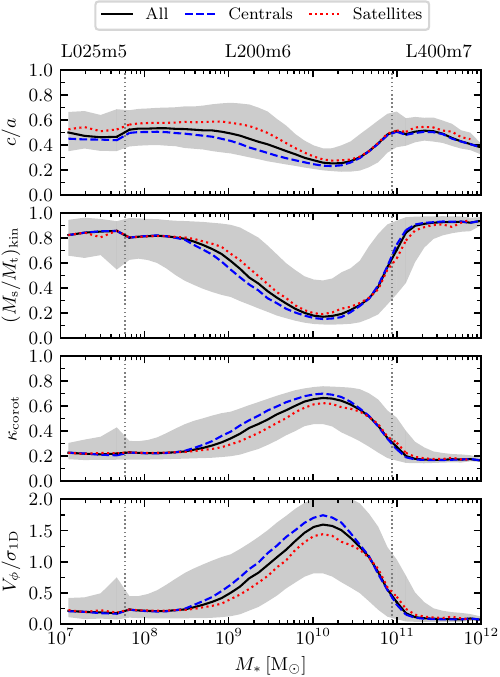}
    \caption{Median galaxy morphology metrics as a function of stellar mass. The shaded regions indicate the 16th to 84th percentiles of the distribution of the whole galaxy population, and the solid line the median values. The {\highres}, {\midres} and {\lowres} simulations are used in different stellar mass ranges, which are delineated by the vertical dashed lines. The galaxy population is split into central (dashed) and satellite (dotted) galaxies. Disc formation in COLIBRE is most efficient for galaxies with $M_{*} \approx10^{10}\,\Msun$. Satellite galaxies are thicker and more spheroid-dominated than central galaxies of similar stellar masses.}
    \label{Figure:correlation_stellar_mass}
\end{figure}

\subsection{The correlation between morphology and stellar mass}\label{Section:CorrelationMorphologyStellarMass}

As we see from Fig.~\ref{Figure:correlation_morphological_metrics}, different mass samples occupy distinct regions of the morphology metric parameter space, suggesting a trend of morphology with stellar mass. Additionally, the extent of the region that each sample occupies varies, indicating that the scatter also depends on the stellar mass of the galaxy. In this section we explore in more detail the median trend with stellar mass, as well as the scatter.

Fig.~\ref{Figure:correlation_stellar_mass} shows the median value of the four morphology metrics as a function of galaxy stellar mass. We show the trend with stellar mass for the whole galaxy population (solid), and for galaxies that are centrals (dashed) or satellites (dotted). To sample the distribution over a wide range of stellar mass, we use each of the three different resolution simulations for a distinct range in stellar mass. We motivate combining the boxes by the good convergence across simulations, as discussed in \S\ref{convergence_simulation_resolution}. By doing so, we can balance the needs for statistical power and numerical resolution, allowing us to robustly sample the predicted morphologies of galaxies over five orders of magnitude in stellar mass. 

We determine which stellar mass range to assign to each simulation by identifying in {\midres} the maximum stellar mass bin that contains an insufficient number of galaxies (poor statistics) and the minimum mass stellar bin that is dominated by poorly-resolved galaxies (numerical effects). We use the {\tt m7} simulation above $M_{*} = 10^{11}\,\Msun$, which is in the well-sampled regime of this simulation ($\geq 10^{4}$ particles per galaxy in {\tt m7}). The $(400 \,\mathrm{Mpc})^{3}$ volume provides a factor of eight more galaxies than the $(200\,\mathrm{Mpc})^{3}$ volume, greatly improving the statistics for the most massive and rarest galaxies (18146 galaxies in total in the range assigned to {\lowres}). To avoid the inclusion of poorly resolved galaxies from {\midres}, we use {\highres} below the stellar mass where the galaxies in the {\tt m6} model are poorly resolved. This corresponds to $\approx 6\times10^{7}\,\Msun$ (see Fig.~\ref{Figure:resolution_convergence}), and contains a sample of 689 galaxies with stellar masses above our lower stellar mass bin limit of $10^{7}\,\Msun$. We thus use the {\midres} simulation between $6\times10^{7}\,\Msun$ and $10^{11}\,\Msun$, which yields 357755 galaxies in total. The stellar mass bins have a width of 0.11 dex for {\lowres} and {\midres}, and 0.19 dex for {\highres} because of its small cosmic volume.

As we see from Fig.~\ref{Figure:correlation_stellar_mass}, using these stellar mass thresholds to combine simulations leads to median curves that have no significant discontinuities. There is however a slight dip in the $c/a$ of galaxies with $M_{*}\approx 10^{11}\,\Msun$, which is caused by the transition from {\midres} to {\lowres}, as {\tt m6} galaxies are less flat than {\tt m7} galaxies at that stellar mass (Fig.~\ref{Figure:resolution_convergence_larger_box}). The scatter shows no notable jumps when switching between {\midres} and {\lowres}, although a clear discontinuity in the scatter is seen when transitioning from {\highres} to {\midres}. This could be because the {\highres} simulation has an unrepresentative scatter due to its small volume.

The thinnest and most rotationally-supported galaxies have a stellar mass of $\approx (1 - 2) \times 10^{10}\,\Msun$. The stellar mass range where more than half of the galaxy population can be classified as discs, e.g. $(M_{\rm s}/M_{\rm t})_{\rm kin} <0.5$, $V_{\phi}/\sigma_{\rm1D}\geq1$ or $\kappa_{\rm corot} \geq 0.4$, extends from $\approx (1 -2)\times10^{9}\,\Msun$ to $(6-7)\times10^{10}\,\Msun$. Beyond this stellar mass range, galaxies become thicker and dispersion-dominated. The transition region between morphologies at the massive end happens relatively fast. Most galaxies are dispersion-dominated above $\approx (1 - 2) \times10^{11}\Msun$, with little scatter in morphology parameters. In contrast, the transition towards dispersion-dominated galaxies at the low-mass end is more gradual. It is only below $\approx (2 - 5)\times10^{8}\,\Msun$ that most galaxies become spheroid dominated.

At the high stellar mass end, the transition between morphological types is relatively well understood \citep[e.g.][]{Oser.2010, Dubois.2013, Clauwens.2018, Tacchella.2019, Park.2022}. Galaxies with stellar masses $M_{*}\sim10^{10}\,\Msun$ are the most efficient at forming stars (see fig. 13 of \citealt{Schaye.2025} for COLIBRE), which preferentially happens in a disc configuration. At higher masses, star formation is quenched (fig. 18 of \citealt{Schaye.2025}) and so disc formation is unable to proceed. The only manner in which galaxies can subsequently grow in stellar mass is therefore by merging with nearby galaxies. This is a disruptive process that changes the phase-space distribution of pre-existing stars, likely leading to a change from rotation- to dispersion-dominated kinematics.

Less well understood is why the lowest mass dwarf galaxies appear to be dominated by spheroids\footnote{In reality, some dwarf galaxies would be classed as irregular rather than spheroids, but none of our chosen morphology parameters measure irregularity. Quantifying galaxy irregularity will be essential for morphology studies at higher redshifts, where irregular galaxies are expected to become more common \citep[][]{Trayford.2019}.}, are consequently thick, and why a wide variety of simulations seem to agree on this \citep[e.g.][]{Tremmel.2017, Clauwens.2018, Tacchella.2019, Zeng.2024, Benavides.2025a}. The first question is whether stars form already resembling a spheroidal morphology, or if the spheroidal morphology results from physical processes capable of changing stellar morphology. For example, potential fluctuations driven by gas outflows can thicken and expand the pre-existing stellar distribution \citep[e.g.][]{Pontzen.2012, El-Badry.2016}, but they can also prevent the settling of a rotation-dominated gas disc \citep[][]{El-Badry.2018}. Successive mergers could also prevent the settling of a gas disc, due to misaligned gas accretion and spin flips of the gas \citep[e.g.][]{Dekel.2020}. Ignoring the effects of feedback and focusing purely on gas physics, the gas disc scale height in dwarf galaxies may become comparable to its radial scale \citep[][]{BenitezLlambay.2018}. This happens because the scale-height of a gas disc depends on the ratio of the gas sound speed to its midplane density, with the difference between a self-gravitating and a non-self-gravitating disc being the power law index. The sound speed is set by the physical conditions in the gas, which for low-mass galaxies is mostly in the warm, atomic interstellar gas phase (e.g. fig. 19 of \citealt[][]{Schaye.2015}).

Both post- and pre-stellar-birth thickening remain plausible explanations, but interpreting the results from previous cosmological simulations is complicated by their over-pressurisation of the interstellar medium (affecting pre-birth properties) and their large dark matter to baryon particle mass ratios (affecting post-birth evolution; see Appendix \ref{Appendix:NumberOfDMParticlesOnMorphology}) and the particular choice of subgrid physics \citep[e.g.][]{Celiz.2025}. Hence, the fact that COLIBRE predicts dispersion-dominated low mass dwarf galaxies warrants further investigation into their formation mechanisms.

Regardless of the true origin of the prevalence of dispersion- and spheroid-dominated dwarf galaxies in simulations, observational data also seemed to indicate that dwarf galaxies are dispersion-dominated and thick \citep[e.g.][]{Wheeler.2017}. Nonetheless, more recent studies suggest that simulated low mass galaxies tend to be too thick \citep[e.g.][]{Xu.2024, Klein.2025}, with some proposing that the stellar mass range where thin discs exist is much lower than previously thought \citep{Benavides.2025}. Understanding whether this claimed tension is present for COLIBRE will require a like-to-like comparison to observational data and a sample selection whose colours and stellar masses are consistent with those found in observations. Selecting a population of galaxies with systematically different star formation activities (most simulated dwarf galaxies are quenched) will change the measured thinness (differences between luminosity- and mass-weighting will increase) and because more actively star forming galaxies have more massive discs (as we will show in \S\ref{Section:CorrelationMorphologyOtherGalaxyProperties}).

Focusing now on satellite galaxies, we see from Fig.~\ref{Figure:correlation_stellar_mass} that their morphologies are systematically different from central galaxies of the same stellar mass. For the kinematic metrics, the morphologies of satellites approach those of the field at the highest and lowest stellar masses, differing only in the stellar mass range where the fraction of disc galaxies becomes non-negligible. The axis ratio $c/a$ is systematically larger, even for the lowest mass satellite galaxies, indicating that they are not as flattened as their central counterparts. These differences could arise naturally through the effects of gravitational tides and interactions with nearby galaxies.

\subsection{The correlation between morphology and intrinsic colour}\label{Section:CorrelationMorphologyStellarMassColour}

\begin{figure}
    \centering
    \includegraphics{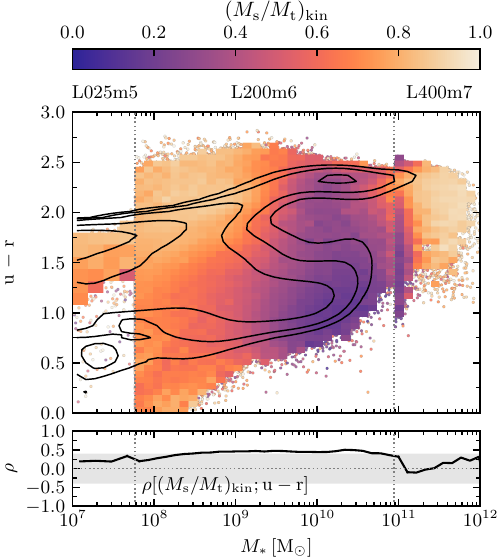}
    \caption{\textit{Top panel}: Distribution of galaxies in the space of intrinsic (u-r) colour and stellar mass, and the correlation with the spheroid-to-total mass ratio. We combine central and satellite galaxies of the {\highres}, {\midres} and {\lowres} simulations in the same manner as in Fig.~\ref{Figure:correlation_stellar_mass}, with the stellar mass range assigned to each simulation indicated in the upper part of the panel and by vertical dashed lines. The bin colour encodes the average spheroid-to-total mass ratio of galaxies, which is only shown if said bin contains more than ten galaxies. In bins with fewer galaxies, each individual galaxy is shown as a dot, with the colour representing its spheroid-to-total mass ratio. Given the relatively small number of galaxies in {\highres}, we use coarser bins than for the {\midres} and {\lowres} simulations. The contours enclose 95, 90, 80, 50 and 25 per cent of the total number of galaxies, after accounting for differences in cosmic volume across simulations and smoothing the counts with a Gaussian kernel with a constant covariance of 0.2. \textit{Bottom panel}: Pearson correlation coefficient between spheroid-to-total mass ratio and galaxy intrinsic (u-r) colour, at a fixed stellar mass. The shaded region indicates values of $\rho$ that correspond to a weak correlation ($|\rho|\leq 0.4$).}
    \label{Figure:colour_vs_stellar_mass_vs_disc_to_total_mass_ratio}
\end{figure}

Observations show that red galaxies tend to be irregular or spheroid dominated, whereas blue galaxies tend to be dominated by a stellar disc component \citep[e.g.][]{Strateva.2001, Baldry.2004, Driver.2006}. We explore whether this is the case in COLIBRE by examining, in Fig.~\ref{Figure:colour_vs_stellar_mass_vs_disc_to_total_mass_ratio}, how intrinsic (i.e. neglecting attenuation by dust) colour\footnote{Defined here as the difference between the total intrinsic u and r band luminosities, expressed in absolute magnitudes, of all bound stellar particles within 50\,kpc from the galaxy centre.}, stellar mass and spheroid-to-total mass ratio relate to each other.

As in the previous subsection, we combine the results of the three different resolutions together. We then bin central and satellite galaxies according to their colour and stellar mass, using a coarser binning for the {\highres} simulation due to its small number of galaxies. For bins that enclose more than ten galaxies, we overlay a pixel whose colour denotes the average $(M_{\rm s}/M_{\rm t})_{\rm kin}$ value of the galaxies therein. For the regions in the colour-stellar mass parameter space without pixels, we show individual galaxies as dots, whose colour is their measured $(M_{\rm s}/M_{\rm t})_{\rm kin}$. To indicate how galaxies are distributed in this space, we overlay contours that enclose several percentiles of the galaxy number count distribution, after smoothing it with a Gaussian kernel with a constant covariance of 0.2 and accounting for differences due to the volume of each simulation.

We see that galaxies occupy a broad range of colours at a fixed stellar mass, with the lowest values of (u-r) becoming less populated by galaxies as the galaxy stellar mass increases. The abrupt change observed at $M_{*}=6\times10^{7}\,\Msun$ is because the number of galaxies sampling the extremes of the colour distribution at a fixed stellar mass is fewer in {\highres} relative to {\midres}, due to the difference in cosmic volume. In contrast, switching from the {\midres} to the {\lowres} simulation does not result in such a drastic change, as the extremes of the colour distribution are well-sampled by both volumes. There is, however, a discontinuity in the average $(M_{\rm s} / M_{\rm t})_{\rm kin}$ of galaxies at fixed u-r colour for (u-r)~$\leq 2$ in the transition from {\midres} to {\lowres}. This happens because galaxies have slightly more massive discs in the {\tt m7} model than in {\tt m6} at this stellar mass (Fig.~\ref{Figure:resolution_convergence_larger_box}), which reflects the fact that quenched fractions differ the most between these resolutions at around this stellar mass (fig. 18 of \citealt{Schaye.2025}). This also causes $M_{*}\approx 10^{11}\,\Msun$ galaxies in {\lowres} to be slightly redder than galaxies in {\midres}.  

Focusing on the overlaid contours, we see that there are two stellar mass ranges where a bimodality in the colours of galaxies exists. One colour bimodality happens for the lowest mass dwarf galaxies in our sample ($M_{*}\approx2\times10^{7}\Msun$). Most galaxies at this stellar mass are red, but there is a small concentration of blue dwarf galaxies at (u-r) $\approx 0.5$. However, because of the aforementioned volume limitations of {\highres}, a larger sample is required to determine whether this blue subpopulation of low-mass dwarf galaxies is robust or driven by noise. 

The second bimodality in colour is most prominent between $M_{*} \approx 3\times 10^{9}\,\Msun$ and $M_{*} \approx 10^{11}\,\Msun$. The contour corresponding to the reddest galaxies in this stellar mass range traces out the red sequence, which spans a relatively narrow range in (u-r) and is flat relative to the red population of galaxies below $M_{*} \approx 3\times 10^{9}$. The bulk of the red galaxy population in this stellar mass range consists of satellite galaxies (not shown). The blue cloud is much broader in its colour distribution at a fixed stellar mass, and contrary to the red sequence, it comprises mostly central galaxies. Some satellite galaxies can be found in the blue cloud, but these are likely recently accreted central galaxies that have not yet undergone environmental quenching. There is a distinct under-density of galaxies between the blue and red population of galaxies, which is the sparsely-populated transitional green valley. 

The bimodality in the colour distribution of galaxies disappears above $M_{*} \approx 10^{11}\,\Msun$, since central galaxies of increasingly large stellar masses become redder and eventually become indistinguishable from the colour distribution characteristic of the red sequence. Consequently, the properties of satellite and central galaxies also become increasingly similar. This evolution in colour as a function of the stellar mass reflects the fact that star formation in the most massive galaxies in COLIBRE is quenched. Below $M_{*} \approx 3\times 10^{9}\,\Msun$, the blue cloud and red sequence become less well defined, as most galaxies span a wide range of colours between $M_{*}\approx 10^{8}\,\Msun$ and $M_{*}\approx 10^{9}\,\Msun$. It is only at $M_{*}\approx 10^{9}\,\Msun$ that two distinct tracks of colour start appearing. 

The only stellar mass range where discs dominate regardless of colour is $(0.6 - 3)\times 10^{10}\,\Msun$. However, as can be seen from the bottom panel of Fig.~\ref{Figure:colour_vs_stellar_mass_vs_disc_to_total_mass_ratio}, there is a mild dependence on colour even at this stellar mass. The Spearman rank correlation coefficient\footnote{This correlation metric relies only on the relative rank ordering of variables (and is thus insensitive to outliers) and, compared to the Pearson correlation coefficient, it is better suited for variable correlations that are non-linear.} between spheroid-to-total mass ratio and colour (at fixed stellar mass) is $\rho \approx  0.5$, meaning that bluer galaxies have more massive disc components than redder galaxies of the same stellar mass. Note that the existence of a non-zero correlation between these two galaxy properties does not require a population of bulge-dominated galaxies in this stellar mass range, as we already showed in Fig.~\ref{Figure:correlation_stellar_mass} and Fig.~\ref{Figure:colour_vs_stellar_mass_vs_disc_to_total_mass_ratio} that galaxies of these stellar masses are generally disc-dominated. The correlation reflects that, within the population of disc-dominated galaxies, the importance of the bulge changes depending on the colour of the galaxy.

We also note the existence of a population of intrinsically red galaxies that are disc-dominated, although a more like-to-like comparison is required to check whether this population resembles real intrinsically red disc galaxies \citep{Mahajan.2020, Cui.2024}. The existence of red discs in COLIBRE suggests that quenching precedes morphological transformation for this population of galaxies, which agrees with simulations \citep[e.g.][]{Martig.2009, Correa.2019, Xu.2022} and observations \citep[e.g.][]{Masters.2010, Bundy.2010, Bell.2012, Gentile.2025}.

Virtually all galaxies with masses above $2\times 10^{11}\,\Msun$ are spheroidal in nature, regardless of their colour. Consequently, the correlation weakens for the most massive galaxies ($\rho \lesssim 0.3$), and actually reaches a minimum of $\rho = 0$ at $M_{*}\approx2 \times10^{11}\,\Msun$. This may seem counterintuitive based on how $(M_{\rm s}/M_{\rm t})_{\rm kin}$ appears to vary with galaxy colour for galaxies around this stellar mass in Fig.~\ref{Figure:colour_vs_stellar_mass_vs_disc_to_total_mass_ratio}. However, the vast majority of galaxies at this stellar mass have (u-r)~$\approx 2.4$, so the apparent dependence of spheroid-to-total mass ratio on colour is driven by outlier galaxies that span a wide range of colours. In fact, the fact that high mass galaxies have (u-r)~$\leq 2$ indicates that they are not all completely quenched in COLIBRE. 

The lowest mass dwarf galaxies ($M_{*}\leq10^{9}\,\Msun$) are also typically spheroid-dominated, but there is still a weak dependence of morphology on colour ($\rho \approx 0.3$), i.e. dwarf galaxies are more likely to contain more massive discs if they are bluer. More massive dwarf galaxies ($10^{9} \leq M_{*}\,[\Msun]\leq10^{10}$) have a stronger correlation between their morphology and colour, at a level similar to the one measured for galaxies with $M_{*}\sim10^{10}\,\Msun$. In fact, blue (u-r~$\leq 1.5$) high-mass dwarf galaxies are generally disc-dominated.

Finally, we note that we have looked at average trends of $(M_{\rm s}/M_{\rm t})_{\rm kin}$ in the colour and stellar mass parameter space. Individual galaxies of a given colour and stellar mass do not necessarily follow the average trend, as can already be seen from the individual galaxy dots in Fig.~\ref{Figure:colour_vs_stellar_mass_vs_disc_to_total_mass_ratio}. In the following two subsections we quantify how the scatter in $(M_{\rm s}/M_{\rm t})_{\rm kin}$ correlates with the properties of galaxies and their host dark matter haloes.

\subsection{The relation between morphology and host halo properties}\label{Section:CorrelationMorphologyHaloProperties}

We have seen that galaxy morphology correlates with stellar mass (\S\ref{Section:CorrelationMorphologyStellarMass}) and colour (\S\ref{Section:CorrelationMorphologyStellarMassColour}). We would now like to quantify how strongly the morphology of a galaxy correlates with its properties and those of its host halo, at a fixed stellar mass. We will do so by using the Spearman rank correlation coefficient between galaxy morphology, here represented by the spheroid-to-total mass ratio $(M_{\rm s}/M_{\rm t})_{\rm kin}$, and other selected halo/galaxy properties. To bin galaxies, we use the same stellar mass range and bins as used in \S\ref{Section:CorrelationMorphologyStellarMass}.

We first explore the correlation between galaxy morphology and the properties of host haloes. The six halo properties that we consider are the virial mass ($M_{\rm 200c}$), concentration ($c_{\rm 200c}$), spin parameter ($\lambda_{\rm 200c}$), sphericity ($c_{\rm halo}/a_{\rm halo}$), triaxiality ($T_{\rm halo}$) and maximum circular velocity ($V_{\rm max}$). The first five are calculated using all particles within $R_{\rm200c}$, as described in \S\ref{Halo_property_calculation}. Because the masses of haloes are dominated by dark matter, these five halo-scale properties will largely reflect the spatial distribution and angular momentum of dark matter. The maximum circular velocity is calculated using only particles that are bound to the central in question, regardless of their particle type. Since spherical overdensities are ill-defined for satellite subhaloes, we only consider central subhaloes in this subsection.

At least three of the halo properties we measure are sensitive to baryonic physics. The concentration and maximum circular velocity of a halo can increase in response to the condensation of baryons at its centre and virial masses decrease due to the feedback-driven loss of baryons \citep[e.g.][]{Sawala.2013, Velliscig.2014, Schaller.2015, Chua.2017, ForouharMoreno.2022}. Hence, to decouple the effect of baryons on the correlations, we measure the halo properties in two ways. The first way is to calculate the properties of haloes in the hydrodynamical simulation, which includes baryonic effects. The second approach to measure the properties of the matched counterparts of the hydrodynamical haloes in the DMO simulation. For this purpose, we do a bijective matching of central subhaloes between the DMO and hydrodynamical simulations (see \S\ref{Section:SubhaloMatching}). Comparing the correlations between morphology and properties of central subhaloes in the DMO and hydrodynamical simulations allows us to gauge the importance of baryonic effects on the host halo properties for the correlations. Since these correlations depend on how well converged morphologies are across different resolutions, as well as the convergence of the host halo properties, we provide convergence tests in Appendix \ref{Appendix:CorrelationConvergence}.

Fig.~\ref{Figure:morphology_correlation_halo_properties} shows the correlation strength between the spheroid-to-total stellar mass ratio and the aforementioned halo properties, measured in the DMO (dashed) and hydrodynamical (solid) simulations as a function of stellar mass. For four halo-scale properties ($M_{\rm 200c}, c_{\rm 200c}, \lambda_{\rm 200c}$, and $c_{\rm halo}/a_{\rm halo}$), the strength of the correlation depends on stellar mass. Halo triaxiality does not correlate with galaxy morphology. The correlation between $V_{\rm max}$ and spheroid-to-total mass ratio has a weak dependence on stellar mass, remaining approximately constant except for the highest mass galaxies.

Focusing on the effect of halo mass, we see that it correlates mildly ($\rho\approx0.4$) with $(M_{\rm s}/M_{\rm t})_{\rm kin}$ for $10^{10}\lesssim M_{*}\,[\Msun] \lesssim10^{11}$. The halo mass is unimportant for more and less massive galaxies, although there is still a very weak correlation ($\rho \approx 0.1$) with the morphology of dwarf galaxies at $M_{*}\approx 2\times 10^{8}\,\Msun$. There is only a very small shift in the correlation strength when considering DMO vs hydrodynamical halo properties. Note that the correlation we observe implies that galaxies with more massive discs reside in haloes in which galaxy formation is more efficient (i.e. larger $M_{*}/M_{\rm 200c}$), which is in agreement with observational inferences \citep[e.g.][]{Wang.2012, Mandelbaum.2016, Posti.2021} and other simulation suites \citep[e.g.][]{Correa.2020, Rodriguez.2025}. 

\begin{figure*}
    \centering
    \includegraphics{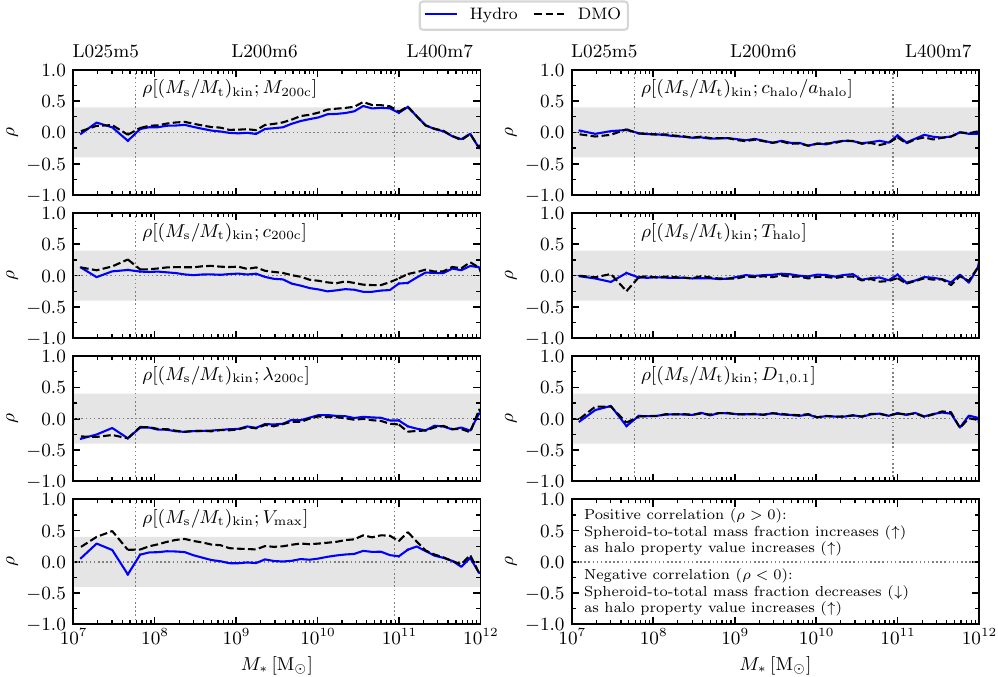}
    \caption{Spearman rank correlation coefficients for the relations between the morphology of central galaxies and various host halo properties: virial mass ($M_{\rm 200c}$), concentration ($c_{\rm 200c}$), spin parameter ($\lambda_{\rm 200c}$), maximum circular velocity ($V_{\rm max}$), sphericity ($c_{\rm halo}/a_{\rm halo}$), triaxiality ($T_{\rm halo}$) and environment ($D_{\rm 1,0.1}$). We show the correlations with halo properties measured in the hydrodynamical (solid) and DMO (dashed) versions of the simulation. The shaded region indicates values of $\rho$ that correspond to a weak correlation ($|\rho|\leq 0.4$). The bottom right panel indicates how different correlation signs reflect how $(M_{\rm s}/M_{\rm t})_{\rm kin}$ changes at fixed stellar mass with a given halo property. The correlation of morphology with halo-scale properties is generally weak and depends on stellar mass.}
    \label{Figure:morphology_correlation_halo_properties}
\end{figure*}

The fact that galaxies at a fixed stellar mass residing in more massive haloes tend to be more spheroid-dominated than galaxies within less massive haloes can be understood based on the fact that halo clustering is mass-dependent \citep[][]{Kaiser.1984}. Galaxies hosted by more massive haloes are generally located in denser environments, which means that they tend to experience a larger number of mergers and interactions than their equal-stellar-mass counterparts in less dense regions \citep[e.g.][]{Fakhouri.2009}. These processes perturb and may even destroy stellar discs, driving a positive correlation between halo mass and spheroid-to-total mass ratio. This may also explain why the importance of halo mass peaks in the stellar mass range where disc galaxies dominate (see Fig.~\ref{Figure:correlation_stellar_mass}). If there is no significantly massive disc to disrupt, then the environment and interactions with neighbouring galaxies will not be a driving mechanism. 

The spheroid-to-total mass ratio anti-correlates with halo concentration in the stellar mass range where disc galaxies dominate, although it is weaker ($\rho \approx-0.3$) than for halo mass. Since halo concentration depends on halo mass (i.e. more massive haloes are generally less concentrated) the correlation between spheroid-to-total mass ratio and halo concentration can be understood from the same reasoning as for the halo mass. Galaxies hosted by less concentrated (more massive) haloes have more massive spheroid components, likely because they have experienced more interactions and mergers. For concentration, we also see more pronounced differences compared to halo mass in the correlation strengths with the halo properties measured in the DMO and hydrodynamical simulations, likely because concentration depends on the scale radius of the halo, whereas halo mass is insensitive to how mass is distributed within $R_{\rm200c}$.

The spin of the halo correlates only weakly with $(M_{\rm s}/M_{\rm t})_{\rm kin}$ in the stellar mass range with a substantial population of spheroid-dominated galaxies, and does not correlate at all in the disc-dominated regime ($10^{10} \leq M_{*} \;[\Msun]\leq 10^{11}$). Among galaxies with $M_{*}\leq 10^{9}\,\Msun$ and $M_{*}\geq 10^{11}\,\Msun$, those residing in higher spin haloes tend to have a more massive disc component than galaxies in lower spin haloes. The overall weak and mass-dependent correlation between morphology and halo spin broadly agrees with other studies using different simulations \citep[e.g.][]{Sales.2012, Rodriguez-Gomez.2017, Rodriguez.2025}. The correlation between halo spin and morphology is essentially the same for the hydrodynamical and DMO properties of the halo, indicating little-to-no change in the population-wide distribution of halo spin. The lack of a baryonic effect on halo spins can be understood from the fact that the spin of a halo is set by tidal torques during its proto-evolutionary growth \citep[e.g.][]{Peebles.1969, Doroshkevich.1970, White.1984} and by the accretion of satellites \citep[e.g.][]{Vitvitska.2002, Maller.2002, Hetznecker.2006}. The inclusion of baryons does not drastically alter either process, and hence the spin remains relatively unaffected by the inclusion of baryons \citep[e.g.][]{Bryan.2013}.

The maximum circular velocity of a halo measured in the hydrodynamical simulation plays a minor role ($\rho \approx0.1$) in the morphology of a galaxy at a fixed stellar mass. Since $V_{\rm max}$ is a proxy for halo mass, it surprising that the correlation strength is much weaker than for $M_{\rm 200c}$, particularly because $V_{\rm max}$ is more relevant on the scale of the galaxy within the halo. Indeed, we see that the largest change in the correlation strength between galaxy morphology and halo properties measured in the hydro and DMO simulations occurs for $V_{\rm max}$, which correlates much more strongly ($\rho = 0.4$) when measured from the DMO halo properties. The fact that the correlation weakens when using the hydrodynamical halo properties may appear, at first instance, counterintuitive. As we will show in \S\ref{Section:CorrelationMorphologyOtherGalaxyProperties}, galaxies of a given stellar mass that have more massive spheroids are more compact. Hence, purely relying on arguments based on the mass distribution of the stars, one might expect that the correlation with $V_{\rm max}$ would be stronger in the hydrodynamical simulation, as more spheroidal galaxies tend to have a more concentrated stellar component. 

To explore why this does not happen, we have verified that the ratio of $V^{\rm hydro}_{\rm max} / V^{\rm DMO}_{\rm max}$ is higher for disc-dominated galaxies than for spheroid-dominated galaxies (not shown). The difference occurs because galaxies with more massive discs contain more baryons (at fixed stellar mass, galaxies with more massive discs contain more gas; see \S\ref{Section:CorrelationMorphologyOtherGalaxyProperties}) and reside in less dark-matter-dominated haloes (at fixed stellar mass, galaxies with more massive discs tend to be hosted in less massive haloes; top panel of Fig.~\ref{Figure:morphology_correlation_halo_properties}). Ultimately, the weakening of the correlation between $V_{\rm max}$ and morphology highlights the fact that $V^{\rm DMO}_{\rm max}$ is a tracer of the past halo assembly history, but $V^{\rm hydro}_{\rm max}$ is not because it also depends on the efficiency with which baryons condense at the centres of haloes.

Out of the two halo shape metrics that we consider, halo triaxiality shows no correlation with the mass fraction of the spheroid component of the galaxy, and the sphericity of the halo only exhibits a weak anti-correlation ($\rho = -0.2$) at $M_{*}=10^{10}\,\Msun$. The anti-correlation between sphericity and spheroid-to-total mass ratio indicates that galaxies with more massive spheroids reside in less-spherical haloes. This reflects the fact that more-massive haloes are less spherical than less-massive ones \citep[e.g.][]{Jeeson-Daniel.2011,Bonamigo.2015}, and that galaxies with more massive spheroids reside in more massive haloes (top panel of Fig.~\ref{Figure:morphology_correlation_halo_properties}).

The correlation strengths between galaxy morphology and either indicator of halo shape are not affected by the inclusion of baryons, implying that the shapes of the haloes within $R_{\rm 200c}$ are not significantly altered by baryons. This does not exclude the (expected) possibility that the shape of the inner halo changes in response to the condensation of baryons, but this effect weakens towards the outskirts of haloes \citep[e.g.][]{Bryan.2013, Chua.2019}.

\begin{figure*}
    \centering
    \includegraphics{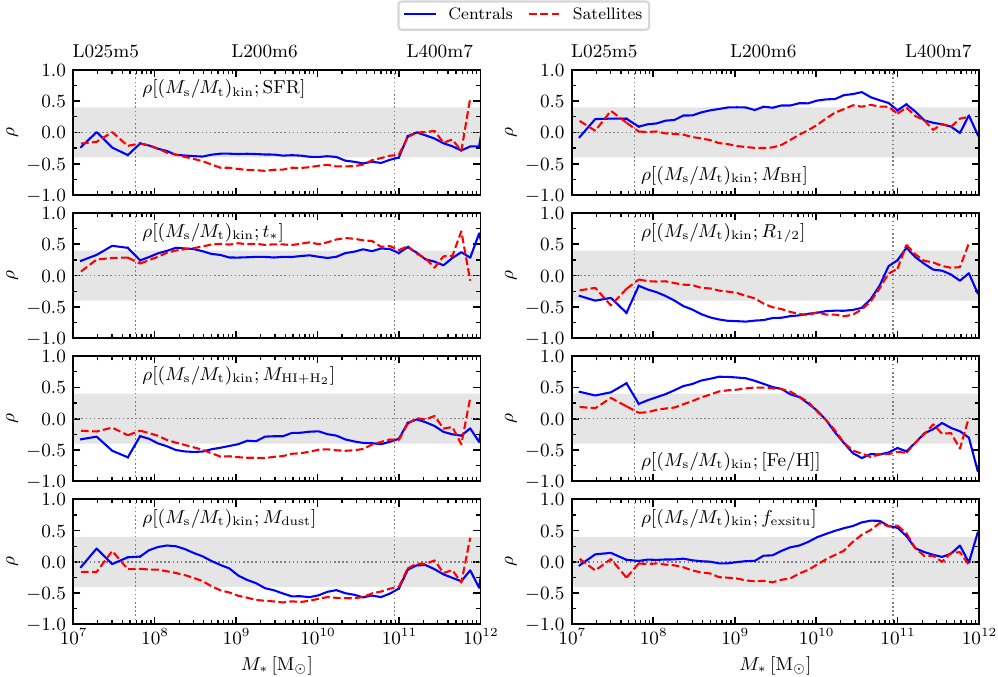}
    \caption{Spearman rank correlation coefficient for the relations between galaxy morphology and various internal properties. The galaxy properties are measured using bound particles within 50~kpc. We show the correlation strength of the central (solid line) and satellite (dashed line) galaxy populations separately. The correlation strengths between morphology and internal galaxy properties are generally strong, mass-dependent and different for central and satellite galaxies.}
    \label{Figure:morphology_correlation_galaxy_properties}
\end{figure*}

Lastly, we see that the spheroid-to-total mass ratio of central galaxies does not correlate with the large-scale environment of the halo,  as quantified by the $D_{\rm 1,0.1}$ measurement defined in section \S\ref{Halo_property_calculation}. This may seem counterintuitive based on the previous discussion relating morphology with halo mass and hence environment, but two important points are worth remembering. First, the environment metric we use has been chosen to be insensitive to halo mass, since environment and halo mass are often difficult to disentangle \citep[][]{Haas.2012}. Second, the median value of $D_{1, 0.1}$ in our sample of haloes is approximately 10 (not shown), so it measures the environment at scales much larger than the halo virial region and of the galaxy itself. Hence, despite the lack of correlation between our environment metric and morphology, our results do not preclude the possibility that more traditional metrics used to quantify environment do correlate significantly with morphology. However, as already noted, such correlations may be a consequence of morphology and environmental metrics both correlating with halo mass.

In summary, at a fixed stellar mass, the spheroid-to-total mass ratio of galaxies correlates only weakly with the properties of their host haloes. Most of the correlations we have seen appear to be reflective of the environment of the galaxy, which has likely influenced its past evolution. Nonetheless, halo properties that correlate weakly when measured on the scale of $R_{\rm 200c}$ may correlate more strongly with galaxy morphology when considering scales more relevant to the galaxy. For example, in the EAGLE simulations, the spatial and angular momentum distribution of stars correlates more strongly with the inner properties of the halo \citep[][]{Zavala.2016, Thob.2019}. 

\subsection{The relation between morphology and internal galaxy properties}\label{Section:CorrelationMorphologyOtherGalaxyProperties}

Having explored the correlation between morphology and halo properties for central galaxies at fixed stellar mass, we now consider the correlation between morphology and a subset of internal galaxy properties, also at fixed stellar mass. We perform the same analysis as in \S\ref{Section:CorrelationMorphologyHaloProperties}, but split the population according to whether they are central or satellite galaxies (we only considered central galaxies in \S\ref{Section:CorrelationMorphologyHaloProperties}). All galaxy properties use a mass-weighted approach, and except for the spheroid-to-total stellar mass ratio and half-stellar-mass radius, are calculated using only bound particles within an aperture of 50~kpc.

Similar to the previous subsection, we provide convergence tests of the correlation strength between $(M_{\rm s} / M_{\rm t})_{\rm kin}$ and the galaxy properties we consider in Appendix \ref{Appendix:CorrelationConvergence}. In summary, the existence of a correlation or anti-correlation is robust across resolutions, so long as there are sufficient particles to resolve the galaxy morphology and the property in question. In some cases, the convergence is better for morphology than for a given galaxy property and in other cases the galaxy property converges better than morphology. To ensure consistency across all panels, we use the same resolution simulations for the same stellar mass ranges as the previous subsections ({\highres} below $M_{*} = 6 \times10^{7}\,\Msun$, {\lowres} above $M_{*} = 10^{11}\,\Msun$ and {\midres} between these two stellar masses).

We show in Fig.~\ref{Figure:morphology_correlation_galaxy_properties}, as a function of stellar mass, the correlation strength of the spheroid-to-total stellar mass ratio with eight galaxy properties: the instantaneous star formation rate (SFR), mean mass-weighted stellar age ($t_{*}$), total atomic and molecular hydrogen mass ($M_{\rm{HI} + \rm{H}_{2}}$), dust mass ($M_{\rm dust}$), mass of the most massive black hole ($M_{\rm BH}$), half-stellar-mass radius ($R_{1/2}$), linearly mass-weighted average stellar metallicity ([Fe/H]) and ex-situ\footnote{Ex-situ stars are those which formed in a galaxy different from the main progenitor of the galaxy they are bound to at $z = 0$. This definition excludes unbound stellar material.} fraction ($f_{\rm exsitu}$). Although $f_{\rm exsitu}$ is not directly observable, it is included in our analysis because we expect it to correlate with galaxy morphology. Note that several of the galaxy properties that we have chosen correlate with each other, e.g. star-forming galaxies tend to contain more gas and dust and less massive black holes than quenched galaxies. Hence, identifying the primary and secondary drivers of galaxy morphology at a fixed stellar mass would require a careful removal of correlations amongst galaxy properties.

Morphology exhibits a correlation with several galaxy properties. At a fixed stellar mass, central and satellite galaxies with a more massive spheroid component tend to contain less gas, have lower star formation rates and older stars, which is in qualitative agreement with observations \citep[e.g.][]{Blanton.2003, Lang.2014, vandeSande.2018, Wang.2020, Wang.2025}. These trends are present for all stellar masses below $M_{*} \approx 10^{11}\,\Msun$, but weaken above this stellar mass. The weakening of these correlations occurs at the mass above which the galaxies quench, most stellar mass growth is driven by mergers and most galaxies become spheroid-dominated. Hence, morphology is driven by external factors that are by definition not considered in our chosen (internal) galaxy properties. An additional factor in the weakening of correlations at high stellar mass is that the presence of a reasonably massive accreted stellar component would likely mask out any pre-existing correlations that exist when considering in-situ stars only.

We also see that the correlation between morphology and the above properties ($\mathrm{SFR}$, $t_{*}$ and $M_{\mathrm{HI}+\mathrm{H}_{2}}$) can become stronger for satellite galaxies than for central galaxies. This difference in correlation strength is maximal in the range $4\times10^{8}\leq M_{*}\,[\Msun]\leq 4\times10^{10}$. This is the regime where environmental effects lead to substantial differences in the satellite and central galaxy population, for example in their colour, resulting from environmental quenching (leading to the two branches seen in Fig.~\ref{Figure:colour_vs_stellar_mass_vs_disc_to_total_mass_ratio}). Hence, environment is likely to be driving the strengthening of these correlations. Understanding why the correlations strengthen for satellite galaxies requires a more detailed analysis. However, we expect that satellites that have lost most of their gas and have become quenched are more likely disturbed and hence have more massive spheroid components. If the timescales for these evolutionary processes are similar, the observed correlation should strengthen for satellites.

The correlations of morphology with other galaxy properties are more sensitive to the stellar mass of the galaxy. The correlation strength with dust mass is similar to the correlation with $M_{\mathrm{HI}+\mathrm{H}_{2}}$, but they deviate below $M_{*}\approx 10^{10}\,\Msun$. This could be explained by the fact that the dust mass is not only related to how much gas a galaxy has, but also to the gas-phase metallicity. In any case, except for the lowest mass dwarf galaxies, more dust-rich galaxies have more massive stellar discs. 

The correlation between morphology of central galaxies and their most massive black hole mass is positive, meaning that more spheroid-dominated galaxies at a fixed stellar mass host more massive black holes. This is consistent with observations, e.g. \citet[][]{Graham.2023}. The correlation strength exhibits a relatively smooth dependence on stellar mass, peaking at $\approx 3\times10^{10}\,\Msun$ and tapering off towards the low and high mass end. The reduced correlation strength with black hole mass at the low-mass end is resilient to resolution (see Fig.~\ref{Figure:morphology_correlation_convergence_centrals}) and likely reflects the fact that black holes are not as important in dwarf galaxies as for more massive ones. The change at the high mass end is likely because, as argued earlier in this subsection, morphology is driven by mergers and because nearly all galaxies become spheroid-dominated. The weakening of the correlation with black hole mass for satellites could be because environmental effects like stripping affect the extended stellar component but not the nuclear black hole, potentially erasing and even reversing (as seen below $M_{*}\approx 7\times10^{9}\,\Msun$) any previously existing correlation. 

The strong anti-correlation between morphology and the half-stellar-mass radius indicates that more extended galaxies have a more prominent disc component, as expected \citep[e.g.][]{Shen.2003,vanderWel.2014b}. For central galaxies, the correlation strength is approximately constant except that it weakens when most galaxies become spheroid-dominated (above $\approx 10^{11}\,\Msun$). The weakening towards lower stellar masses is likely driven by numerical effects, as we see a strong resolution dependence on the measured correlation between morphology and galaxy size (see Fig.~\ref{Figure:morphology_correlation_convergence_centrals}). 

The correlation with stellar metallicity shows the strongest mass dependence out of all the properties discussed here. Above $M_{*} \approx 10^{10} \Msun$, metallicity anti-correlates with the spheroid-to-total stellar mass ratio, meaning that more metal-rich galaxies have less massive spheroid components. Below $10^{10}\,\Msun$ the trend reverses, and more metal rich galaxies have a more massive spheroid relative to their disc. The stellar mass dependence of the correlation strength between morphology and stellar metallicity in COLIBRE is similar to the correlation between gas-phase metallicity and morphology found for EAGLE galaxies \citep[][]{Zenocratti.2020}. We interpret this trend with stellar mass to be primarily driven by an increasing fraction of accreted material with increasing galaxy mass. At low stellar masses, the ex-situ fraction is low, and the correlation between stellar metallicity and spheroid mass fraction reflects the fact that galaxies whose spheroid is more massive tend to be more compact (and are hence more metal rich, e.g. \citealt[][]{Scott.2017}). This is also in line with the fact that more concentrated galaxies (i.e. with more massive bulges) are capable of retaining more of their metals and hence have higher metallicities than galaxies where discs are more massive \citep[][]{Wang.2026}. Galaxies that have a large fraction of accreted stellar mass naturally contain a more massive spheroid (i.e. stellar halo or intra-cluster mass) which is itself more metal-poor than the in-situ component. The accreted component has a lower metallicity than the in-situ stars because it is built up from the disruption of less massive galaxies, which are more metal-poor as per the mass-metallicity relation. Since the fraction of accreted stellar mass increases with the stellar mass of the galaxy, this effect drives the observed correlation at high stellar masses. 

The last correlation we consider is with the ex-situ stellar mass fraction of galaxies (bottom right panel of Fig.~\ref{Figure:morphology_correlation_galaxy_properties}). At a fixed stellar mass, the correlation with ex-situ fraction for central galaxies is positive above $2\times10^{9}\,\Msun$ and peaks at $M_{*}\approx5\times10^{10}\,\Msun$. The stellar mass at which the correlation strength peaks is similar to what was found for IllustrisTNG \citep[][]{Tacchella.2019}, TNG50 \citep[][]{Park.2022} and EAGLE \citep[][]{Proctor.2024}. Galaxies with larger ex-situ mass fractions have more massive spheroid components, which is expected because the accreted stellar component is part of our spheroid definition. For satellite galaxies, we actually observe an anti-correlation below $2\times10^{10}\,\Msun$, so that satellites with larger accreted mass fractions are more disc-dominated. This difference in qualitative behaviour from the relation observed in central galaxies is likely due to environmental processing, like the removal of the weakly bound accreted component due to gravitational tides.

\section{Conclusions}\label{Section:conclusions}

We presented the first overview of the stellar morphologies of present-day galaxies formed in the COLIBRE suite of simulations. The COLIBRE simulations are particularly well-suited for morphology studies owing to several factors. First, COLIBRE removes the artificial pressure floor that is present in most galaxy formation models that simulate representative samples of the universe. The presence of this so-called `equation of state' in previous simulations may have caused gas discs to become over-pressurised, which would eventually propagate to the spatial and kinematic distributions of the star particles that they form. Second, dark matter particles in COLIBRE are super-sampled relative to baryonic particles, which suppresses the spurious evolution of galaxy morphologies \citep[e.g.][]{Wilkinson.2023} and results in galaxies with smaller minor-to-major axis ratios at the same baryonic particle mass (Appendix \ref{Appendix:NumberOfDMParticlesOnMorphology}). Lastly, COLIBRE has been shown to reproduce a variety of observable scaling relations (e.g. \citealt[][]{Schaye.2025,Chaikin.2025b,Lagos.2025, Ludlow.2026}).
Given all of the above, and the fact that morphologies were not directly considered during the calibration of the COLIBRE model, we set out to explore galaxy morphologies in COLIBRE.  

In this work, we adopted a theory-space approach to quantify the stellar morphologies of galaxies. This means that we have full information about which star particles belong to which galaxies, regardless of how close any two resolved galaxies may be. We additionally have full knowledge of the stellar phase-space distribution. This choice allows us to explore in detail the actual morphologies of galaxies, although it makes comparisons to observations difficult, which we defer to future work.

We carefully explored the effect of several choices required to quantify the morphologies of galaxies. We found that:
\begin{itemize}
    \item Using iterative inertia tensors reduces the bias introduced by a spherical aperture. Compared to a non-iterative approach, the iterative inertia tensors detect significantly larger fractions of flattened and disc galaxies, regardless of the chosen aperture size (Fig. \ref{Figure:inertia_tensor_joint}).
    \item The aperture size used to measure the morphology of galaxies influences the resulting values. By systematically varying the aperture used to measure the inertia tensor (Fig.~\ref{Figure:inertia_tensor_joint}), we found that too small an aperture underestimates the contribution of the disc. On the other hand, using a very large aperture makes the inertia tensor sensitive to structure external to the galaxy itself, like its stellar halo. We find that an aperture radius of three times the half-stellar-mass radius is a good compromise between these extremes, with similar conclusions found when looking at kinematic morphology indicators (Fig.~\ref{Figure:median_shift_metrics_aperture_effect}). 
    \item Using luminosity-weighted morphology metrics biases results relative to a mass-weighted approach. Despite good angular momentum alignment between these two choices (Fig.~\ref{Figure:angular_momentum_misalignment}), mass-weighted quantities result in less rotationally-dominated and thicker galaxies than using even the reddest photometric GAMA band (K; Fig.~\ref{Figure:luminosity_weighting}). Overall, luminosity-weighting can extend the stellar mass range where disc galaxies dominate down by an order of magnitude. 
\end{itemize}

Having established a fiducial method with which to measure the morphology of galaxies, which relies on mass-weighted quantities measured within $3R_{1/2}$, we explored the convergence of, and correlation between our chosen parameters. 

\begin{itemize}
    \item The number of stellar particles below which the galaxy morphology becomes dominated by shot noise is small. By subsampling galaxies and comparing them to the ground truth, we found that only $\approx 50$ particles are required to recover median morphology trends for both for kinematic and spatial metrics (Fig.~\ref{Figure:sampling_particle_convergence}). When shot noise dominates, galaxies become more rotationally-dominated and flatter than the true population. This means that morphology measurements are less sensitive to shot noise for disc galaxies than for elliptical galaxies.  
    
    \item Based on the largest volume available at $z = 0$ for all three COLIBRE resolutions, we find good convergence in morphological trends with stellar mass (Fig.~\ref{Figure:resolution_convergence}). There is a shift in the morphology-stellar mass relation between the {\tt m7} and {\tt m6} models, but the differences are typically small, $\Delta(M_{\rm s}/M_{\rm t})_{\rm kin} \leq 0.05$ (Fig.~\ref{Figure:resolution_convergence_larger_box}).

    \item The different morphology indicators correlate very well with each other, particularly the kinematic measures (Fig.~\ref{Figure:correlation_morphological_metrics}). Despite the strong correlations between the different parameters, there is some amount of scatter, partly caused by differences in definitions and partly by other galaxy properties we have not considered in our analysis. This means that results based on investigations of galaxy samples selected by a cut in a single morphology parameter should be interpreted with caution.
\end{itemize}

We explored how galaxy morphology relates to the stellar mass, intrinsic colour and environment of a galaxy. We combined the three largest available boxes for each COLIBRE resolution to sample galaxies across five orders of magnitude in stellar mass.

\begin{itemize}
    \item Galaxies with $M_{*}\approx (1-2)\times10^{10}\,\Msun$ are the most rotation-dominated and flattest (Fig.~\ref{Figure:correlation_stellar_mass}). Between $M_{*}\approx 10^{9}\,\Msun$ and $7\times10^{10}\,\Msun$, the majority of galaxies are disc-dominated. Dispersion-dominated galaxies only become ubiquitous above $2 \times10^{11}\Msun$ and below $5\times10^{8}\,\Msun$.
    \item Depending on the stellar mass of the galaxy, morphology may correlate with colour more or less strongly (Fig.~\ref{Figure:colour_vs_stellar_mass_vs_disc_to_total_mass_ratio}). High-mass dwarf galaxies ($10^{9}\leq M_{*}\,[\Msun] \leq 4\times10^{9}$) are disc-dominated if they are blue, but spheroid-dominated if they are red. Galaxies with stellar masses near $10^{11}\,\Msun$ show a similar trend, but the transition between the disc- and spheroid-dominated regime occurs at redder colours than in the high-mass dwarf galaxy population. Galaxies with stellar masses in the range where the disc population dominates are almost always disc-dominated regardless of their colour (e.g. there are red discs), although bluer galaxies have more massive discs relative to redder galaxies. The highest mass galaxies ($\geq2\times10^{11}\,\Msun$) are spheroid-dominated regardless of their colour.
\end{itemize}

Lastly, we quantified the correlation strength between galaxy morphology and several host halo and galaxy properties at fixed stellar mass. To gauge by how much the inclusion of baryons affects the correlation strength between galaxy morphology and host halo properties, we used the properties of haloes measured in the hydrodynamical simulation and of the matched counterparts in the DMO simulations.

\begin{itemize}
    \item The correlation between morphology and halo properties is generally weak and stellar-mass-dependent (Fig.~\ref{Figure:morphology_correlation_halo_properties}). The concentration ($c_{\rm 200c}$), virial mass ($M_{\rm 200c}$) and maximum circular velocity ($V_{\rm max}$) of the host halo play a role in the stellar mass range where disc galaxies dominate, with halo mass being the property that correlates the strongest with morphology. Galaxies at a fixed stellar mass that reside in haloes that are more massive, less concentrated and have higher maximum circular velocities contain more massive spheroid components. In contrast, halo spin ($\lambda_{\rm 200c}$) correlates weakly in the stellar mass range where spheroidal galaxies dominate: higher spin haloes contain galaxies that are more disky. The connection between the shape of the halo and galaxy morphology is also weak, but more spherical haloes contain galaxies with more massive discs. Halo triaxiality does not correlate with morphology.   
    
    \item The correlations between galaxy morphology and galaxy properties that quantify their gas content and star formation activity are strong (Fig.~\ref{Figure:morphology_correlation_galaxy_properties}). At a fixed stellar mass, galaxies with more gas, higher star formation rates and younger stellar populations have more massive disc components. A similar trend is also seen for the dust mass, whereby more dusty galaxies have more prominent discs, but this trend weakens below stellar masses of $10^{9}\,\Msun$. The different dependence on the morphology-dust correlation with stellar mass compared to the morphology-gas correlation likely reflects the fact that additional processes determine the dust mass of a galaxy (e.g. gas metallicity).

    \item The correlation of morphology with other galaxy properties we have investigated generally shows a stronger stellar mass dependence than with gas content and star formation rate (Fig.~\ref{Figure:morphology_correlation_galaxy_properties}). Galaxies with more massive black holes tend to be more spheroidal, with the correlation strength peaking at $M_{*}\approx 4\times10^{10}\,\Msun$ and decreasing towards lower stellar masses. This stellar mass dependence likely reflects the reduced influence of black holes on the evolution of dwarf galaxies. The stellar mass dependence is the strongest between morphology and stellar metallicity, which is correlated with the spheroid-to-total mass ratio below $M_{*}\approx 10^{10}\,\Msun$ and anti-correlated above this stellar mass. At high-mass, we attribute this trend to an increase of the fraction of accreted stellar mass with increasing stellar mass. Galaxies with a larger ex-situ stellar fraction contain relatively more metal-poor stars and have more spheroidal morphologies following the mergers that build up the accreted component. We indeed see an increasing importance of the ex-situ fraction in determining the morphology of the host galaxy as its mass increases.

    \item The strength of the correlations between morphology-SFR, morphology-gas content and morphology-stellar age increases for satellite galaxies, particularly in the stellar mass range where environmental processing plays an important role (Fig.~\ref{Figure:morphology_correlation_galaxy_properties}). Conversely, the correlation strengths for the morphology-galaxy size and morphology-metallicity weaken for satellite galaxies compared to centrals. For certain correlations, such as those for the morphology-black hole mass and morphology-accreted stellar mass fraction relations, satellite galaxies exhibit the opposite trend than the one measured for centrals. 
    
    \item All the correlations between morphology and other galaxy properties weaken for $M_{*}\geq 10^{11}\,\Msun$  (Fig.~\ref{Figure:morphology_correlation_galaxy_properties}). This reflects the fact that the highest-mass galaxies have similar (spheroidal) morphologies regardless of their internal properties and, additionally, that factors beyond  internal galaxy properties could be driving the morphology of the stellar component. The reduced diversity of morphology for low-mass dwarf galaxies ($M_{*}\lesssim 5\times10^{7}\,\Msun$) may also play a role in weakening morphology correlations with galaxy properties, but the correlations do not weaken as much as for high-mass galaxies.
    
\end{itemize}

In summary, COLIBRE exhibits a diversity of galaxy morphologies. Galaxy morphology depends on stellar mass, with discs dominating at $M_{*}\approx(1 - 2)\times 10^{10}\,\Msun$. Morphology correlates weakly with the properties of the host halo at fixed stellar mass, but does so more strongly with internal galaxy properties, particularly for those that characterise its star formation activity and gas content (e.g. stellar age, SFR, total atomic and molecular hydrogen mass). We have characterised what effect certain operational choices (e.g. aperture size, mass- vs luminosity-weighting) have on the measured morphologies of galaxies, finding that they have a non-negligible effect. This fact underscores the importance of using forward-modelling when comparing with observations, which we plan to do in a future study. Thanks to their combination of relatively high-resolution, large volumes and physical realism, the COLIBRE simulations provide a unique opportunity to investigate the origin of the morphology of galaxies and its connection with their environment and past evolutionary history.

\section*{Acknowledgements}

We thank the anonymous referee for their helpful comments.

This work used the DiRAC@Durham facility managed by the Institute for Computational Cosmology on behalf of the STFC DiRAC HPC Facility (www.dirac.ac.uk). The equipment was funded by BEIS capital funding via STFC capital grants ST/K00042X/1, ST/P002293/1, ST/R002371/1 and ST/S002502/1, Durham University and STFC operations grant ST/R000832/1. DiRAC is part of the National e-Infrastructure. VJFM acknowledges support by NWO through the Dark Universe Science Collaboration (OCENW.XL21.XL21.025). SP acknowledges support by the Austrian Science Fund (FWF) through grant-DOI: 10.55776/V982. ABL acknowledges support by the Italian Ministry for Universities (MUR) program “Dipartimenti di Eccellenza 2023-2027” within the Centro Bicocca di Cosmologia Quantitativa (BiCoQ), and support by UNIMIB’s Fondo Di Ateneo Quota
Competitiva (project 2024-ATEQC-0050). JT acknowledges support of a STFC Early Stage Research and Development grant (ST/X004651/1). EC acknowledges support from STFC consolidated grant ST/X001075/1. FH acknowledges funding from the Netherlands Organization for Scientific Research (NWO) through research programme Athena 184.034.002. ADL acknowledges financial support from the Matariki Network of Universities.

\section*{Data Availability}

The data used in this study can be made available upon reasonable request to the corresponding author.


\bibliographystyle{mnras}
\bibliography{references}

\appendix

\section{Hybrid versus thermal AGN feedback}\label{Appendix:hybrid_vs_thermal_morphologies}

\begin{figure}
    \centering
    \includegraphics{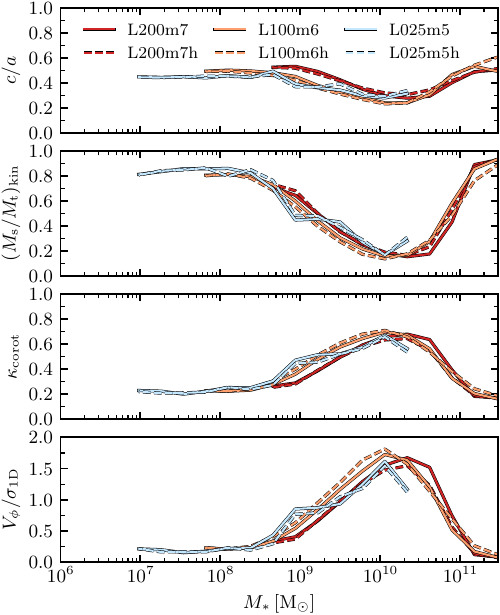}
    \caption{Comparison of how median galaxy morphology varies between simulations with thermal (solid) and hybrid (dashed) AGN feedback models. We use central and satellite galaxies taken from the largest volumes available to both thermal and hybrid at $z=0$, which are $(200\,\mathrm{Mpc})^{3}$ for {\tt m7}, $(100\,\mathrm{Mpc})^{3}$ for {\tt m6} and $(25\,\mathrm{Mpc})^{3}$ for {\tt m5} resolution, respectively. We only show lines for stellar mass bins that contain at least 10 galaxies and for stellar masses that are not shot-noise-dominated. The morphologies of galaxies are very similar regardless of whether the thermal or hybrid AGN feedback model is used.}
    \label{Figure:hybrid_vs_thermal_morphology}
\end{figure}

There are two different implementations of AGN feedback in COLIBRE: one that injects only thermal energy, and another one that injects thermal and kinetic energy. The results we present in our main analysis only use the simulations with purely thermal AGN feedback. Although the two types of models were calibrated to the same set of observables, the simulations with hybrid AGN feedback may differ in the galaxy morphologies that they predict. We check whether this is the case in Fig.~\ref{Figure:hybrid_vs_thermal_morphology}, where we show the median morphology of all galaxies extracted from the largest volumes for which both thermal and hybrid simulations are available at $z = 0$. Differences in galaxy morphology are minimal and typically smaller than the differences between different numerical resolutions (at fixed AGN feedback mode). Some level of difference exists in the stellar mass range where disc galaxies are common ($10^{9}\lesssim M_{*}\,[\mathrm{M}_{\odot}]\lesssim 10^{11}$), which is likely attributable to the somewhat different quenched fractions in the thermal and hybrid simulations (see fig.~25 of \citealt[][]{Schaye.2025}).

\section{The effect of number of dark matter particles on galaxy morphology}\label{Appendix:NumberOfDMParticlesOnMorphology}

One of the distinguishing features of the COLIBRE simulations is that the initial conditions contain four dark matter particles per baryonic particle. The super-sampling of dark matter particles is done so that the DM-to-baryon particle mass ratio is close to unity, which suppresses numerical effects that spuriously affect galaxy morphology \citep[e.g.][]{Wilkinson.2023}. In this appendix, we explicitly test how super-sampling the DM particles propagates to the measured morphologies. 

\begin{figure}
    \centering
    \includegraphics{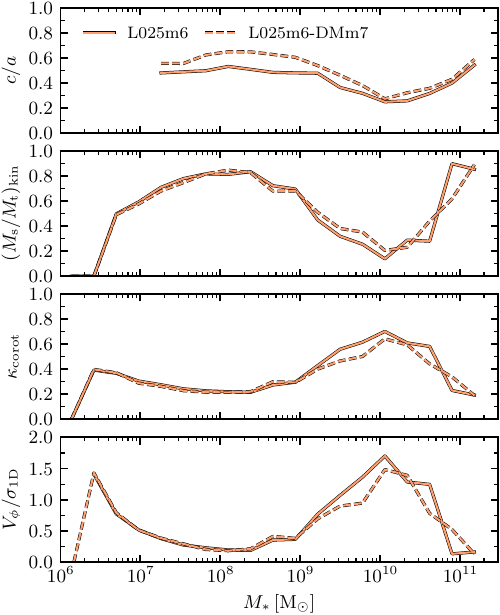}
    \caption{The effect on galaxy morphology of reducing the number of DM particles at a fixed number of baryonic particles in {\tt L025m6} simulations. The solid lines show the median value of each morphology indicator as a function of the stellar mass of the galaxy for the fiducial run, in which there are four DM particles per baryonic particle. The dashed lines show the median values for a run with the same phases and baryonic particle mass in the initial conditions, but using only a single DM particle per baryonic particle. The larger DM-to-baryon particle mass ratio thickens galaxies across all stellar masses, and reduces rotational support and disc importance for galaxies with $10^{9}\leq M_{*}\,[\Msun]\leq 2\times10^{10}$. Note that the inertia tensor calculations require at least 20 stellar particles within the initial aperture, which is why the $c/a$ curves start at higher stellar masses than the kinematic metrics.}
    \label{Figure:effect_equal_DM_particles}
\end{figure}

We run a $(25\,\mathrm{Mpc})^{3}$ box with the same subgrid model and phases in the initial conditions as {\tt L025m6}, but only generating a single dark matter particle per baryonic particle. The median morphology metrics as a function of stellar mass for this modified run are compared to the fiducial {\tt L025m6} run in Fig.~\ref{Figure:effect_equal_DM_particles}. All galaxies are thicker in the simulation with fewer dark matter particles, regardless of whether they are disc- or bulge-dominated or if they are well resolved. The kinematics of the stars in bulge-dominated galaxies are unaffected, but disc-dominated galaxies have lower rotational support when using fewer dark matter particles.

\section{Convergence of morphology correlations with galaxy properties}\label{Appendix:CorrelationConvergence}

In this appendix we discuss how the correlation strength between spheroid-to-total mass ratio and host halo or internal galaxy properties converges with the simulation resolution. To do so, we perform the same analysis as in \S\ref{Section:CorrelationMorphologyOtherGalaxyProperties}, but we use the $(25\,\mathrm{Mpc})^{3}$ volume and measure the correlation strength for the {\tt m7}, {\tt m6} and {\tt m5} resolutions.

\begin{figure}
    \centering
    \includegraphics{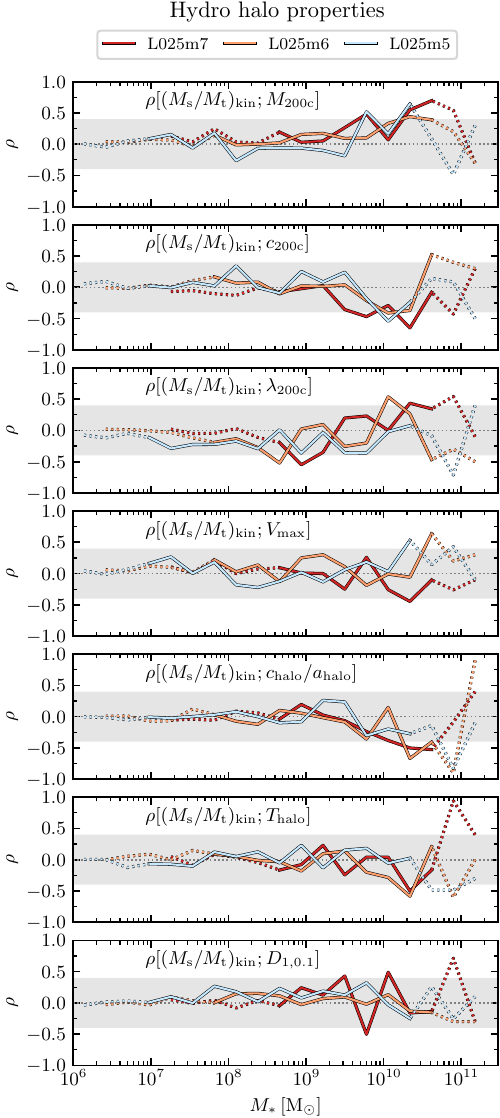}
    \caption{Exploring the numerical convergence of the correlation between spheroid-to-total mass ratio and host halo properties as a function of stellar mass. We only consider $z = 0$ central galaxies taken from {\tt L025m7}, {\tt L025m6} and {\tt L025m5} simulations and that have a bijective match to a central subhalo in the DMO simulation. The dotted lines at the high mass end indicate stellar mass bins that contain fewer than 10 galaxies. The stellar mass bins in which galaxies are poorly sampled ($\lesssim50$ stellar particles) are also indicated by dotted lines. The presence of correlation or anti-correlation is well-converged across different resolutions as long as the number of particles is sufficiently high.}
    
    \label{Figure:correlation_halo_properties_convergence_hydro}
\end{figure}

\begin{figure}
    \centering
    \includegraphics{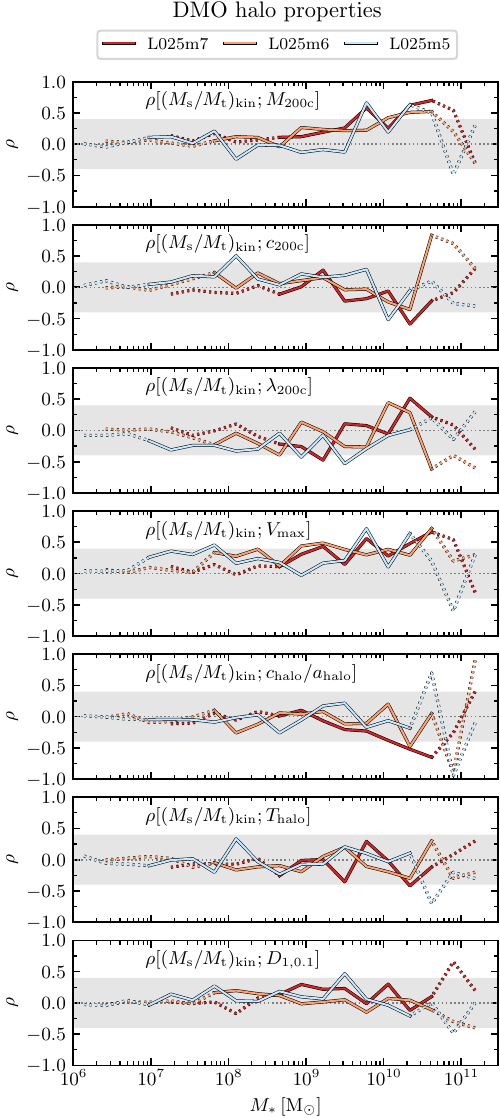}
    \caption{Same as Fig.~\ref{Figure:correlation_halo_properties_convergence_hydro}, but using host halo properties of the matched halo counterparts in the DMO simulations.}
    \label{Figure:correlation_halo_properties_convergence_DMO}
\end{figure}

The correlation strengths between morphology and host halo properties are shown in Fig.~\ref{Figure:correlation_halo_properties_convergence_hydro} (hydrodynamical properties) and Fig.~\ref{Figure:correlation_halo_properties_convergence_DMO} (DMO properties). The existence of a (anti-)correlation is generally robust across resolution levels, as long as the number of particles remains sufficiently high and the correlation is sufficiently strong. For correlations that are particularly weak, e.g. with $V_{\rm max}$ measured in the hydrodynamical simulation, simulations may disagree about the sign of the correlation, but they all nonetheless result in weak correlations.

\begin{figure*}
    \centering
    \includegraphics{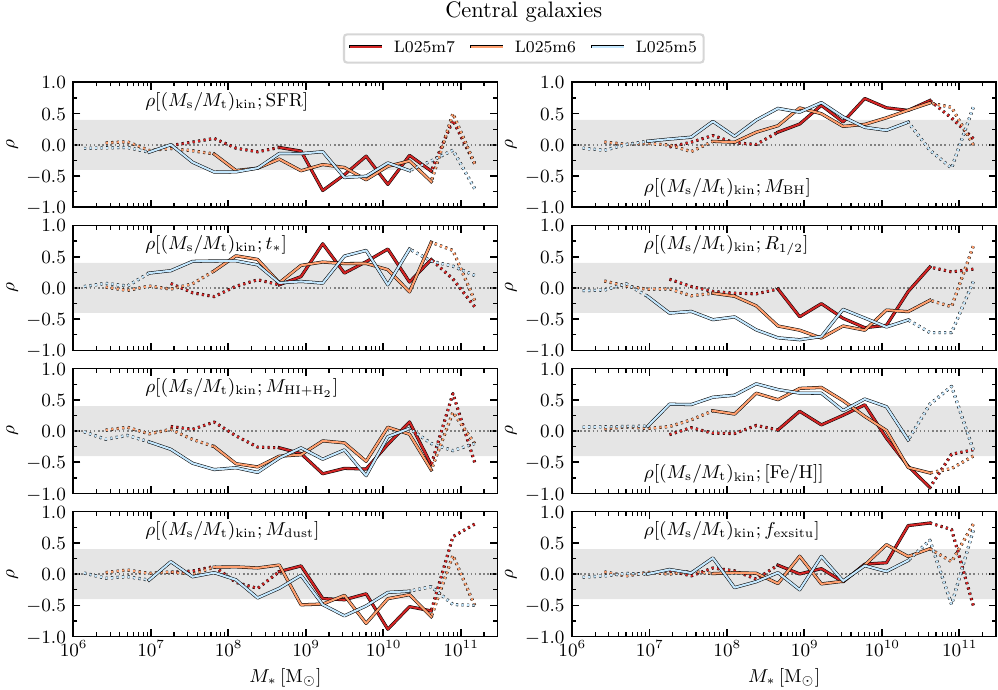}
    \caption{Exploring the numerical convergence of the correlation between spheroid-to-total mass ratio and internal galaxy properties as a function of mass. We only consider $z = 0$ central galaxies taken from {\tt L025m7}, {\tt L025m6} and {\tt L025m5} simulations. The dotted lines at the high mass end indicate stellar mass bins that contain fewer than 10 galaxies. The stellar mass bins in which galaxies are poorly sampled ($\lesssim50$ stellar particles) are also indicated by dotted lines.}
    \label{Figure:morphology_correlation_convergence_centrals}
\end{figure*}

The correlation strength for central galaxies is shown in Fig.~\ref{Figure:morphology_correlation_convergence_centrals}. The existence of a (anti-)correlation is generally robust across resolution levels, as long as the number of particles remains sufficiently high. The magnitude of the correlation may change between resolutions.

Once the number of particles becomes insufficient, the measured correlation weakens until it is dominated by noise, at which point there is no remaining correlation signal. The stellar mass at which this happens is sensitive to how well converged the morphologies of galaxies are at that stellar mass, as well as how well converged a given galaxy property is.

Similar conclusions can be derived for the correlations present in the satellite galaxy population, as seen from Fig.~\ref{Figure:morphology_correlation_convergence_centrals}. Interestingly, certain galaxy properties related to the gas content of galaxies (SFR, $M_{\mathrm{HI}+\mathrm{H}_{2}}$ and $M_{\rm dust }$) show a better agreement between the different resolutions than those obtained for the central galaxy population.

\clearpage

\begin{figure*}
    \centering
    \includegraphics{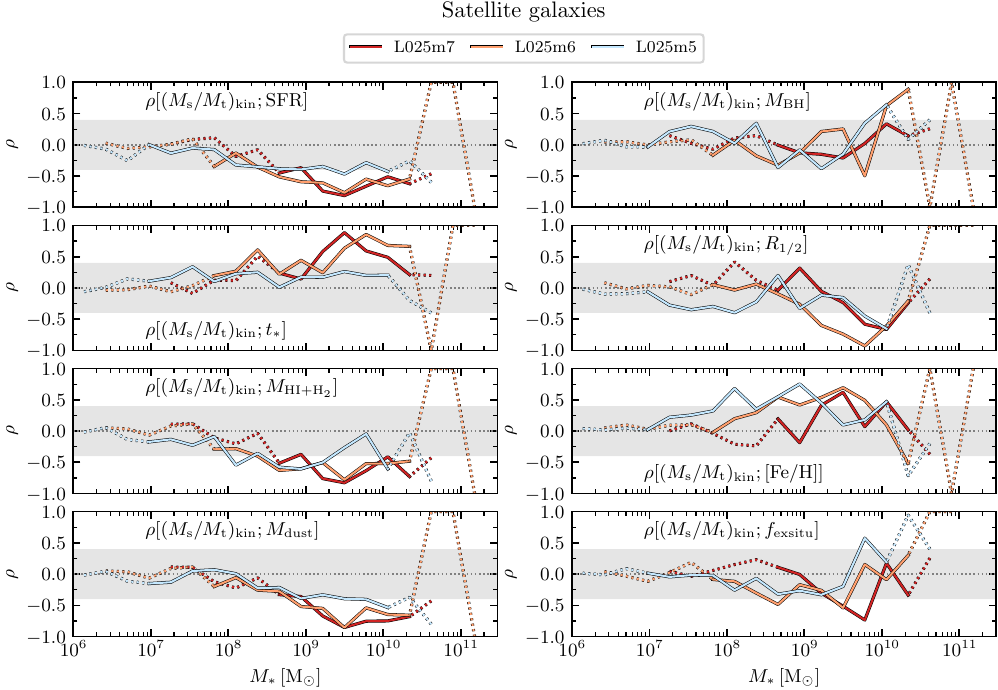}
    \caption{Same as Fig.~\ref{Figure:morphology_correlation_convergence_centrals}, but for satellite galaxies.}
    \label{Figure:morphology_correlation_convergence_satellites}
\end{figure*}

\label{lastpage}
\end{document}